\documentclass[lettersize,journal]{IEEEtran}
\usepackage{amsmath,amsfonts}
\usepackage{algorithmic}
\usepackage{algorithm}
\usepackage{array}
\usepackage[caption=false,font=normalsize,labelfont=sf,textfont=sf]{subfig}
\usepackage{textcomp}
\usepackage{stfloats}
\usepackage{url}
\usepackage{verbatim}
\usepackage{graphicx}
\usepackage{cite}
\begin{document} 

\title{Spatial Mode Multiplexing for  Fiber-Coupled IM/DD Optical Wireless Links with Misalignment%Optimizing SMM-Enabled Indoor FMF Coupling Systems for Real-World Applications: Analyzing of IM/DD Receivers with Stochastic Vibrations
}

\author{Jinzhe Che, Shenjie Huang~\IEEEmembership{Member,~IEEE,}, Majid Safari~\IEEEmembership{Senior Member,~IEEE,}
        % <-this % stops a space
\thanks{The authors are with the Institute for Imaging, Data and Communications,
The University of Edinburgh, Edinburgh EH9 3JL, U.K. (e-mail:
J.Che@sms.ed.ac.uk).}% <-this % stops a space
}
%\thanks{Manuscript received April 19, 2021; revised August 16, 2021.}

\IEEEpubid{}

\maketitle

\begin{abstract}
Optical wireless communication (OWC) emerges as a pivotal solution for achieving terabit-level aggregate throughput in next-generation wireless networks. With the mature high-speed transceivers and advanced (de)multiplexing techniques designed for fiber optics, fiber-coupled OWC can be seamlessly integrated into existing ultra-high-speed networks such as data centres. In particular, OWC leveraging spatial mode multiplexing (SMM) and few-mode fiber (FMF) coupling can significantly increase capacity, though misalignment may reduce performance. This paper presents a thorough investigation into the SMM-enabled FMF coupling OWC systems affected by link misalignment, specifically focusing on systems with intensity modulation with direct detection (IM/DD) receivers. A theoretical analysis is conducted to assess the fiber coupling efficiency of the considered system in the presence of both pointing error and angle of arrival (AOA) fluctuations caused by random device vibrations. Our model elucidates the dependence of coupling efficiency to the order of the incident modes, highlighting the critical role of beam properties in system performance. To mitigate the intermodal crosstalk arising from link misalignment, we employ zero-forcing beamforming (ZFBF) to enhance the overall aggregated data rate.
Through extensive numerical results,
we identify optimal system configurations encompassing aperture design and mode selection, leading to a capacity boost exceeding 200\%.

\end{abstract}

\begin{IEEEkeywords}
Optical wireless communication, spatial mode multiplexing, link misalignment, data center network, few-mode fiber, fiber coupling.
\end{IEEEkeywords}
\section{Introduction}

\IEEEPARstart{T}{he} prevalent radio frequency (RF) based wireless networks face significant throughput limitations due to the increasingly congested RF spectrum, which hinders the support for data-intensive applications like 4K and 8K ultra-high-definition (UHD) video streaming \cite{OWCOverall}. Optical wireless communication (OWC), leveraging the broad spectrum of light waves, presents a promising solution to overcome this spectrum bottleneck.  Additionally, OWC offers heightened security levels as light waves cannot penetrate solid objects\cite{OWCsecurity}. As a pivotal technology within OWC, fiber-coupled OWC systems are gaining increasing recognition for their potential in high-speed data transmission \cite{FiberCouplingDemo3,FiberCouplingDemo2,FiberCouplingDemo1}. This rising prominence can be attributed to their potential for seamless integration with mature high-speed devices designed for fiber optics, including high-fidelity optical signal amplifiers \cite{FiberCouplingDemo3}, precise wavelength filters \cite{FiberCouplingDemo2}, and high-bandwidth fast switching\cite{OWDCNfeasibily}. This integration is crucial for incorporating OWC technology into the existing data communication infrastructures, such as optical data center networks (ODCN) and passive optical networks (PON) \cite{OWCDCN,OWDCNSurvey,PON,OWPON}.

The recent advancement in mode (de)multiplexing technology facilitates efficient integration of the few-mode fiber (FMF) or the multi-mode fiber (MMF) with the prevailing single-mode fiber (SMF) systems \cite{photoniclanternoverview, TaperedPhotonicLattern,ULIPhotonicLattern,PhotonicIntegratedCombiner}. FMF, with its larger core size, can accommodate several orthogonal higher-order modes beyond the fundamental mode supported by SMF systems, thereby alleviating the alignment issue in OWC systems. This capability not only increases fiber coupling efficiency in conditions of turbulence or jitter but also enables efficient mode division multiplexing \cite{TurblenceFMF1, FMFCouplingDis}. In parallel, spatial mode multiplexing (SMM) in free space complements FMF-coupled OWC systems, allowing for simultaneous transmission of multiple data streams through orthogonal mode sets. The comparison of information capacity between SMM systems and traditional multiple-input-multiple-output (MIMO) systems has been discussed in \cite{CapacityOAMMIMO}. It is concluded that SMM performs as the best technique in realizing the capacity limit with the same space–bandwidth product. Specifically, attributing to the smaller space-bandwidth product and simpler multiplexing techniques, the orbital angular momentum (OAM) modes have garnered significant attention, despite constituting only a subset of the complete beam basis \cite{ShenjieSMMDiversity,OAM1,OAMDis}. 

While SMM has the theoretical potential to increase aggregated channel capacity, the performance of SMM-based OWC is significantly hindered by unique OWC channel impairments, especially the misalignment loss. Generally, two types of channel random variation can be introduced by the misalignment issue, namely, pointing error and angle of arrival (AOA) fluctuations.
Pointing error is predominantly attributed to transmitter vibrations, which are commonly introduced by the building sway and the device orientation for outdoor and indoor OWC applications, respectively \cite{Buildingsway, Datacentervibration}. The effect of pointing error can be modeled as the radial displacement of the beam spot position at the receiver\cite{PointingError,FMFCouplingDis}. On the other hand, changes in the beam's spatial phase, driven by turbulence or receiver vibrations, can lead to AOA fluctuations. Such angular deviation can lead to the shift of the diffracted pattern on the focusing optics focal plane \cite{AOA}, resulting in a reduction of the detected power due to the limited size of the fiber core, i.e., coupling loss. In practical applications, the above-mentioned channel impairments induced by misalignments are always coexistent\cite{Datacentervibration}, and their impacts on the intermodal crosstalk and power penalty for the SMM system with free-space photodetector (PD) receivers have been investigated experimentally \cite{OAMDis, LGBEAMDis}. 

To mitigate the crosstalk introduced by mode coupling, traditional fiber optics systems often resort to coherent detection-based MIMO digital signal processing (MIMO-DSP) techniques \cite{MIMOfIBER}, a practice that extends to the design of receivers in OWC systems  \cite{MIMODSPFreeSpace1, MIMODSPFreeSpace2, CapacityOAMMIMO}. Despite their effectiveness, the inherent complexity of MIMO-enabled coherent receivers impedes their large-scale implementation in commercial networks. Consequently, intensity modulation with direction detection (IM/DD) receivers, known for their stability and cost-effectiveness, have been predominantly employed in practical applications \cite{OWCOverall}. In such IM/DD-based multiplexing systems, some works have assumed that the multiple channels are mutually incoherent, thus simplifying the channel representation and enabling the traditional interference mitigation method like singular value decomposition (SVD) \cite{MutuallyIncoherentSVD}. However, this assumption of mutually incoherent channels may be only achieved at the expense of losing spectral efficiency, where employing multiple laser sources at different wavelengths or a single laser with a significantly wide linewidth can diminish the beat noise generated at the output of the square-law PD \cite{MutuallyInCoandCo, MutuallyInDiffLaser}. In this paper, we consider the more practical and spectrally efficient design, where multiple beams are generated by a single or multiple narrow linewidth laser(s) as well as a spatial mode multiplexer \cite{ShenjieSMMDiversity, MutuallyCohernet}. This implies that the interfering optical beams form a coherent superposition at the receiver, i.e., mutually coherent channels. Since the signals are mixed non-linearly in such IM/DD systems, most conventional MIMO schemes are not applicable, but techniques such as zero-forcing beamforming (ZFBF) and optical phase retrieval can be employed to mitigate the intermodal crosstalk \cite{ShenjieSMMZFBF,Phaseretravel1}.

This paper provides a comprehensive theoretical framework to evaluate the communication performance of the SMM-enabled FMF coupling system in the presence of both pointing error and AOA fluctuations. The main contributions of this paper can be summarized as follows:

\begin{itemize}
\item[$\bullet$] We introduce a framework to analyze the performance of an FMF-coupled OWC system characterized by mutually coherent multiplexing channels. By considering the quadratic nature of the IM/DD receivers and misalignment-induced modal crosstalk, we derive the non-linear channel transformation matrix for the proposed system. This addresses a notable research gap, shifting the focus from coupling efficiency models under ideal channel conditions to a comprehensive evaluation of system performance impaired by channel misalignments \cite{FMFCouplingDis, OptimizingFMFEfficiency}.
\end{itemize}

\begin{itemize}
\item[$\bullet$] We propose an accurate theoretical model for indoor multi-mode fiber-coupling channels, offering valuable insights into the joint effects of the pointing error and AOA fluctuations. This model deviates from conventional approaches that rely heavily on the far-field assumption and the transmission of the fundamental Gaussian beam.  Furthermore, we provide a theoretical coupling model for evaluating the coupling efficiency of the FMF across various incident beam modes, facilitating a deeper understanding of system behavior under a broader range of operational scenarios.
\end{itemize}

\begin{itemize}
\item[$\bullet$]  We employ a ZFBF scheme to mitigate the mode crosstalk and analytically derive the aggregated data rate to assess the communication performance of the considered mutually coherent system. We also present some simulation results to provide insights into both the efficiency and communication performance of FMF-coupled OWC under various conditions, indicating the variation of the coupling performance for different mode groups and the significant capacity enhancement introduced by the ZFBF.
\end{itemize}

This paper is organized as follows: Section \ref{2} introduces the system schematic with the channel matrix. Section \ref{3} focuses on modeling fiber-coupled channels and analytically deriving the aggregated data rate equations. Section \ref{SimulationResults} presents some simulation results and discusses the optimal system geometrical configuration. Finally, Section \ref{Conclusion} concludes this paper.

\section{Spatial mode multiplexing system with mutually coherent channels}
\label{2}

\subsection{System Model}
\begin{figure*}[t]
   \centering
   \includegraphics[scale=0.6]{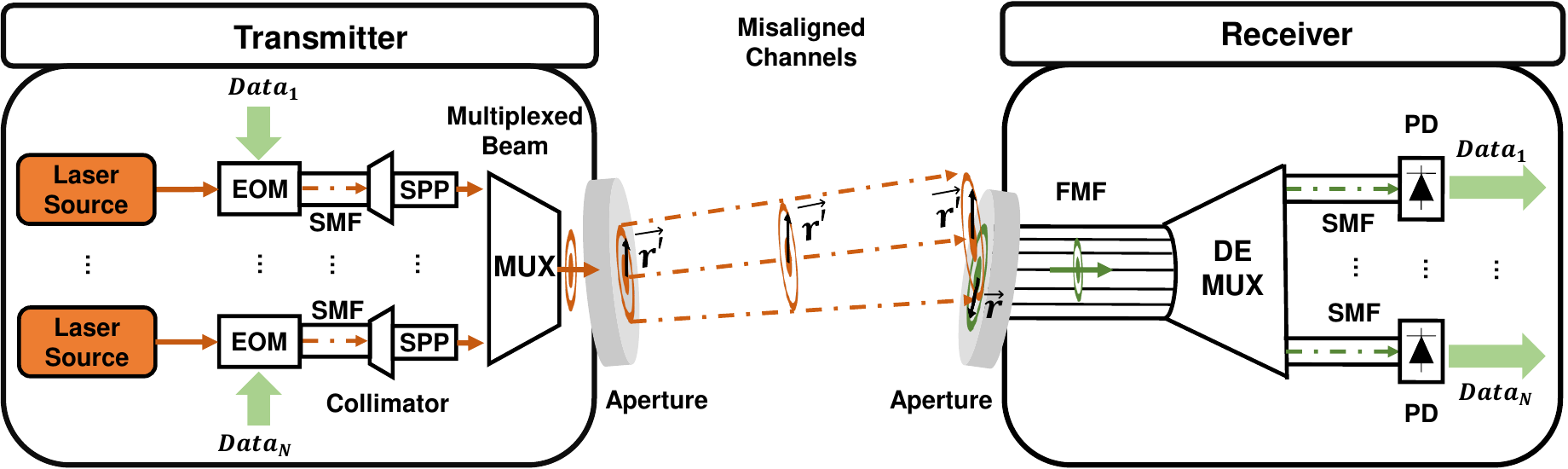}
    \caption{The considered FMF-coupled spatial mode multiplexing OWC system with mutually coherent channels: EOM: electro-optic modulators; SMF: single-mode fiber; SPP: spiral phase
    plate; (DE)MUX: modal (de)multiplexing;  PD: photodetector.  }
    \label{SMM System}
\end{figure*}
Fig. \ref{SMM System} presents the schematic diagram of the proposed SMM-based FMF-coupled OWC system in the presence of link misalignment. For the considered indoor scenarios, the misalignment is normally introduced by the device vibration, which results in the orientation angle deviation for both transmitters and receivers. At the transmitter, the single-wavelength laser sources with a narrow linewidth are employed to realize the mutually coherent system, where all the modulated signals are generated at the same wavelength, thereby realizing a spectral-efficient system design. Electro-optic modulators (EOMs) modulate $N$ input data streams onto the separated collimated Gaussian beams coupled into $N$ SMFs. These modulated beams are then transformed into $N$ orthogonal spatial modes by the spiral phase plates (SPP), with the composite multiplexed beam subsequently propagating over an indoor optical wireless channel suffering from misalignment issues. At the receiver, the incoming multiplexed optical beam, upon its arrival, is initially captured by the aperture and promptly coupled into an FMF. Notably, the pointing error caused by the transmitter vibration introduces the random spatial shift between the beam center and the receiver aperture center. On the other hand, the receiver vibration results in the unparalleled wavefront plane with respect to the slant aperture, leading to the deviation in beam phase distribution over the receiver telescope and the displacement of the diffracted pattern at the FMF fiber end-face (AOA fluctuations). The combined effects of the aforementioned impairments disrupt the orthogonality among free-space channels, and this disruption leads to significant intermodal crosstalk in the light coupled into the FMF. Subsequent to the fiber coupling, the adjacent demultiplexing operation is employed to separate the different spatial modes for isolating each mode before conversion back to the fundamental fiber modes, which are then channelled into respective SMFs tailored for photodetection. Mode-selective photonic lanterns are promising components that can be utilized in the proposed system to realize the efficient transition from the FMF to SMFs with minimal cost and power loss \cite{photoniclanternoverview}. There are two prevalent types of photonic lanterns: tapered and integrated. Tapered photonic lanterns combine several SMFs into a multi-mode core, encapsulated by a cladding of low-index medium to facilitate multi-mode propagation \cite{TaperedPhotonicLattern}. Conversely, integrated photonic lanterns are fabricated by directly etching waveguides onto a chip, enabling advanced stability for splitting light modes in sophisticated devices, albeit with implications such as the increased cost \cite{ULIPhotonicLattern}. Finally, an array of $N$ photodetectors is deployed at the SMF ends to effectively detect the optical power of each spatial mode, which are the superposition of the signal and the interference from other modes generated due to channel impairments.

\subsection{Multi-Mode Optical IM/DD Channel Model }
In the proposed model, the field distribution of the $i$th spatial mode at the transmitter side is denoted as $u'_i(\Vec{r'},0)$, where $\vec{r'}$ indicates the radial vector in the transmitter cylindrical coordinate system as shown in Fig. 1. The beam propagated for a distance $z$ through the misaligned link, denoted by $u'_i(\Vec{r'},z)$, can be projected into the receiver aperture plane defined by the radial vector $\Vec{r}$ as shown in Fig. 1. The projected field, denoted by $\hat{u}_i(\Vec{r},z)$, can be then decomposed into an orthogonal set of fiber modes, backpropagated to the receiver aperture plane, i.e., $u_k(\Vec{r},z)$, given by  \cite{ShenjieSMMDiversity}:
\begin{equation}
    \hat{u}_i(\Vec{r},z) = \sum_{k=-\infty}^{\infty}\mu_{ik}u_k(\Vec{r},z),
\end{equation}
where $u_k(\Vec{r},z)$ denotes the field distribution for $k$th fiber mode back propagated to the receiver aperture plane, which satisfies the orthogonality condition as
\begin{equation}
\int u_k(\vec{r}, z) u_{k^{\prime}}^*(\vec{r}, z) \mathrm{d} \vec{r}= \begin{cases}1, & \text { if } k=k^{\prime} \\ 0, & \text { if } k \neq k^{\prime}\end{cases},
\end{equation}
and $\mu_{ik}$ denotes the channel coefficient for transmitting the $i$th mode and receiving the $k$th  mode, which can be expressed as \cite{ShenjieSMMDiversity}
\begin{equation}
    \mu_{ik} = \int_\mathcal{A} \hat{u}_i(\Vec{r},z)u^{*}_k(\Vec{r},z)\mathrm{d} \vec{r}.\label{Eq.ChannelElement}
\end{equation}
where $\mathcal{A}$ represents the two-dimensional space defined by the receiver aperture.
Assuming the use of an ideal demultiplexer, for the FMF-coupled SMM system with the transmitted mode set $\mathcal{N}$, the total received field at $k$th single-mode fiber can be expressed as
\begin{equation}
     U_{k}(\Vec{r},z,t) =\sum_{i\in \mathcal{N}}\mu_{ik}u_k(\Vec{r},z)s_i(t)e^{j\left(\nu t+\phi_i\right)},
\end{equation}
where $s_i(t)$ denotes the modulated optical amplitude for the $i$th transmitted mode, $\nu$ refers to the center frequency, and $\phi_i$ is the phase difference between the output of the $i$th laser source and the 1st laser source. Note that all lasers are assumed to be narrow linewidth generating the same wavelength of light. In addition, if a single laser is used to generate all transmitted modes, then $\phi_i=0$. Finally, the received current at the square-law PD used to receive the signal of the $k$th spatial mode can be represented as
\begin{align}
        Y_k  & = R\displaystyle \int \left | U_{k}(\Vec{r},z,t) \right |^2\mathrm{d}\Vec{r}\nonumber
        \\
         & = R\displaystyle \int \left | \sum_{i\in \mathcal{N}}\mu_{ik}u_k(\Vec{r},z)s_i(t)e^{j\left(\nu t+\phi_i\right)} \right |^2\mathrm{d}\Vec{r} \nonumber
        \\
         & =  R\left | \sum_{i\in \mathcal{N}}\mu_{ik}s_i(t)e^{j\phi_i} \right |^2,
        \label{Eq.RxCurrentCoh}
\end{align}
with $R$ is the PD responsivity.

Notably, \eqref{Eq.RxCurrentCoh} indicates the coherent superimposition for the intended signal $\mu_{kk}s_k(t)$ and the interference from other subchannels $\mu_{ik}s_i(k)$. Note that the considered coherent superposition differs from the incoherent superposition arising from assuming mutually incoherent channels, which result in a simple intensity superposition as\cite{ShenjieSMMZFBF}:
\begin{equation}
     Y'_k = R\sum_{i\in \mathcal{N}}\left | \mu_{ik}s_i(t) \right |^2.
\end{equation}
However, as  elucidated in the literature, such mutually incoherent framework tends to overly complicate the transmitter design and would lead to the lower spectral efficiency \cite{MutuallyCohernet, ShenjieSMMDiversity, ShenjieSMMZFBF, Phaseretravel1}. Consequently, the following analysis will focus on mutually coherent channels, as defined in (\ref{Eq.RxCurrentCoh}). Therefore, The channel model for the considered mutually coherent SMM systems can be expressed in the following matrix form 
\begin{equation}\label{Eq.ChannelTrans}
    \mathbf{Y} = R\left| \mathbf{H}\mathbf{s}\right|^2+\mathbf{Z},
\end{equation}
where $\mathbf{Y}=[Y_1, \dots, Y_N]^T$ is the vector denoting the received current with totally $N$ spatial modes in set $\mathcal{N}$, $\mathbf{s} = [s_1, ... ,s_N]^{T}$ indicates the signal amplitude vector, and $\mathbf{Z} = [Z_1,Z_2, ... ,Z_N]^{T}$ denotes the noise vector. Normally three main types of noise can affect the performance: thermal noise, shot noise, and relative intensity noise. Considering the thermal noise as the dominant noise source \cite{FOVDataRateTradeoff}, elements in $\mathbf{Z}$ can be assumed to follow a Gaussian distribution with zero mean and a variance of $\sigma^2_n$. Note that $\mathbf{H}$ refers to the complex channel matrix, given by 
\begin{align}
\mathbf{H}&=\left[\begin{array}{ccc}
h_{11} & \ldots & h_{N 1} \\
\vdots & \ddots & \vdots \\
h_{1 N}  & \ldots & h_{N N}
\end{array}\right] \nonumber
\\
&= \left[\begin{array}{ccc}
\mu_{11}e^{j\phi_1} & \ldots & \mu_{N 1}e^{j\phi_N}  \\
\vdots & \ddots & \vdots \\
\mu_{1 N}e^{j\phi_1}  & \ldots & \mu_{N N}e^{j\phi_N} 
\end{array}\right],
\end{align}
where the phase difference between the $\mu_{ik}$ and $h_{ik}$ stems from the considered multiple laser sources and this relationship can be simplified to $\mu_{ik} = h_{ik}$ for the single laser systems. 

\section{Fiber Coupling in the Presence of Misalignment}
\label{3}
The communication performance of the proposed system is closely tied to the characteristics of the channel matrix $\mathbf{H}$, which is directly determined by the coupling efficiency of the considered fiber-coupled system. With the perfectly aligned communication links, the orthogonality among spatial modes can be well preserved, leading to the diagonalization of $\mathbf{H}$. However, in the presence of misalignment, considerable crosstalk between subchannels are introduced, thereby impairing the system stability \cite{OAMDis}. In order to provide insights into the communication performance, the fiber coupling performance in the presence of the misalignment will be comprehensively investigated in this section. In addition, to improve the aggregated capacity, ZFBF is employed at the transmitter to mitigate the interference between mutually coherent subchannels as conventional linear MIMO techniques at the receiver side are not applicable. 
\begin{figure}[t]
   \centering
    \includegraphics[scale=0.4]{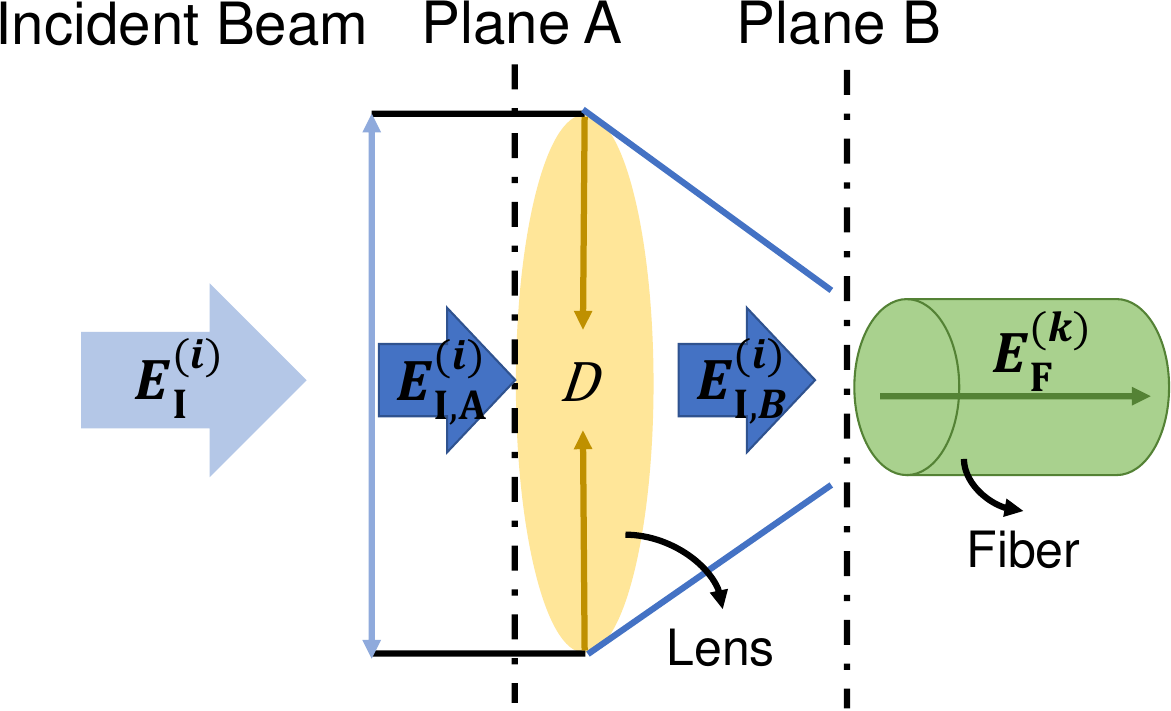}
    \captionsetup{justification=centering}
    \caption{The lens with diameter $D$ couples $i$th incident beam $ {E^{(i)}_{\mathrm{I}}}$ to the $k$th fiber mode field $ {E^{(k)}_{\mathrm{F}}}$. Plane A: aperture; plane B: fiber end face.}
    \label{OpticalCoupling}
\end{figure}

The magnitude of the complex channel coefficient $h_{ik}$ in $\mathbf{H}$ can be estimated by measuring ${P^{(ik)}_{\,\mathrm{F}}}$, which denotes the received optical power associated with the $k$th spatial mode, or equivalently the power coupled into the $k$th mode of the fiber, with the condition that only the $i$th mode is being transmitted with unit power. As illustrated in Fig. \ref{OpticalCoupling}, let's denote ${E^{(i)}_{\mathrm{I,A}}}$ and ${E^{(i)}_{\mathrm{I,B}}}$ as the electric fields of the 
$i$th incident beam at the aperture plane (referred to as plane A) and the fiber plane (referred to as plane B), respectively. 
This fiber-coupled power ${P^{(ik)}_{\,\mathrm{F}}}$ can be then expressed as \cite{SMFCouplingEfficiency1}
\begin{equation}
  \left|h_{ik}\right|^2 = {P^{(ik)}_{\,\mathrm{F}}} = {\left|\int_{\mathcal{B}} {E^{(i)}_{\mathrm{I,B}}} \left ( \vec{r}  \right ) E^{(k)*}_{\mathrm{F,B}}\left ( \vec{r}  \right ) \mathrm{d} \vec{r} \right|^2},
  \label{Eq.CouplingPower1}
\end{equation}
where $E^{(k)*}_{\mathrm{F,B}}$ represent the complex conjugate of the $k$th fiber mode at plane B. As it is more convenient to evaluate the overlap integration on plane A, after yielding the back-propagated fiber mode $E^{(k)}_\mathrm{F,A}$, \eqref{Eq.CouplingPower1} can be rewritten as \cite{SMFCouplingEfficiency2}
\begin{align}
  \left|h_{ik}\right|^2 = \left| \int_{\mathcal{A}}  {E^{(i)}_{\mathrm{I,A}}}(\vec{r}) E^{(k)*}_{\mathrm{F,A}}(\vec{r})\mathrm{d}\vec{r}\right|^2.
 \label{Eq.CouplingPower2}
\end{align}
In addition to determine the magnitude of $h_{ik}$, its phase information can also be acquired through different techniques, such as the method introduced in \cite{ShenjieSMMZFBF}.  

\subsection{Fiber Mode Approximation}

In a weakly guided step-index FMF, the distribution of the guided mode fields $E^{(k)}_{\mathrm{F,B}}$ aligns with the linearly polarized (LP) modes, which are derived from the solutions to the scalar Helmholtz equation. However, the substantial mathematical complexity makes it challenging to accurately assess the coupling loss directly from the LP modes, and previous works tend to adopt Laguerre-Gaussian (LG) modes due to the aligned cylindrical symmetry \cite{FMFCouplingDis}. Derived from the paraxial Helmholtz wave equation in cylindrical coordinates, the LG modes provide a comprehensive representation of the field distribution following free-space propagation. Specifically, at the beam waist position, the existing literature has identified that the overlap relation $M_{lp}$ between $\mathrm{LP}_{l,p}$ and $\mathrm{LG}_{p-1,l}$ can reach 0.99 for the FMF under the low radial order modes, where $M_{lp} = 0$ indicates the orthogonal mode fields and $M_{lp} = 1$ indicates the perfectly matched fields \cite{moderelation}. Thus, to simplify the following derivations, we can utilize LG modes to accurately represent the 
$k$th fiber modes $E^{(k)}_{\mathrm{F,B}}$. To this end, after applying the back-propagation theory, the fiber field over the receiver aperture is given by 
\begin{align}
    &E^{(k)}_{\mathrm{F,A}}\left(\Vec{r}\right) = \mathrm{LG}_{p_\mathrm{F},l_\mathrm{F}}\left(r,\theta\right) = \nonumber
    \\&\frac{B^\mathrm{F}
    _{pl}}{\omega}\left(\sqrt{2}\frac{r}{\omega}\right)^{\left | l_\mathrm{F} \right |} L^{\left | l_\mathrm{F} \right |} _{p_\mathrm{F}}\left(\frac{2r^2}{\omega^2}\right)\exp{\left(-\frac{r^2}{\omega^2}\right)}e^{-j l_\mathrm{F} \theta}.
    \label{Eq.EstimatedFiberModes}
\end{align}
In this context, $p_\mathrm{F}$ and $l_\mathrm{F}$ indicate the radial index and azimuth index of the fiber mode, $L^{\left | l_\mathrm{F} \right |} _{p_\mathrm{F}}$ are the associated Laguerre polynomials, and $B^{\mathrm{F}}_{pl} =\left(\frac{2 p_\mathrm{F} !}{\pi(\left | l_\mathrm{F} \right |+_\mathrm{F}p) !}\right)^{\frac{1}{2}}$ is a normalization factor. The term $\omega = \frac{\lambda f}{\pi \omega_a}$ is the back-propagated mode radius \cite{FMFCouplingDis}, where $\lambda$ is the wavelength, $f$ is the aperture focal length, and $\omega_{a}$ is the fiber mode field radius. This fiber mode field, along with the incident beam field (as derived later in Section \ref{misalignment}), can be substituted into \eqref{Eq.CouplingPower2} for the coupling power.

\subsection{Modeling the Incident Misaligned Beams}\label{misalignment}
\begin{figure}[t]
   \centering
    \includegraphics[scale=0.68]{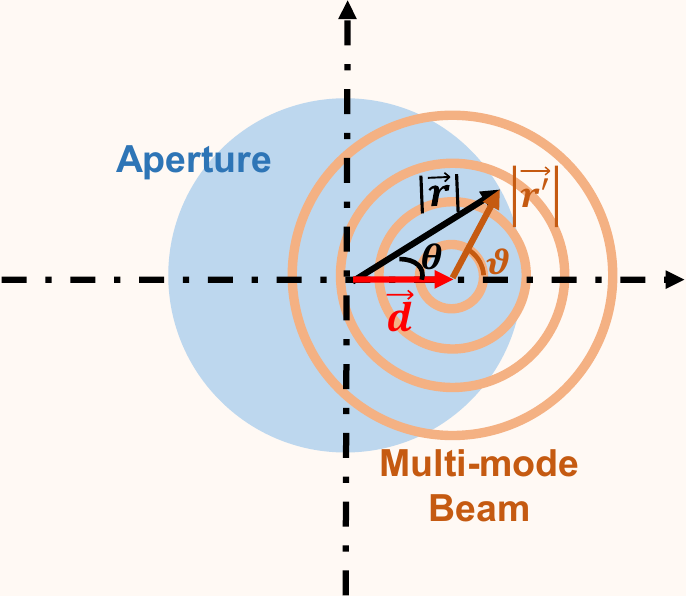}
     \caption{Illustration of the transmitter vibration-induced pointing error, showing the displacement $\Vec{d}$ between the beam center and the aperture center. }
     \label{Displacement1}
\end{figure}
Benefiting from the smaller space-bandwidth product and simpler multiplexing techniques, the OAM multiplexed systems, which use a subset of LG modes, have been notably highlighted in recent works \cite{CapacityOAMMIMO,ShenjieSMMDiversity}. Moreover, LG modes feature better compatibility with many optical subsystems due to their intrinsic circular symmetry, especially for the considered fiber coupling systems \cite{OAM1,OAMDis}. Therefore, in the following discussion, we focus on the transmission of LG modes. Note that the analysis presented here can be readily expanded to the transmission of other spatial modes. 
Therefore, after propagating over a distance $z$ within perfectly aligned channels, the resultant field distribution at the receiver side can be formally articulated as \cite{PhotonicsBook}
\begin{equation}
    \begin{split}
        &{E^{(i)}_{\mathrm{I}}}\left(\Vec{r'},z\right) ={E^{(i)}_{\mathrm{I,A}}}\left(\Vec{r},z\right) = \mathrm{LG}_{p_\mathrm{I},l_\mathrm{I}}\left(r,\theta,z\right) = 
        \\
        &\frac{B^\mathrm{I}_{pl}}{\omega_{z}}\left(\sqrt{2}\frac{r}{\omega_{z}}\right)^{\left | l_\mathrm{I} \right |} L^{\left | l_\mathrm{I} \right |} _{p_\mathrm{I}}\left(\frac{2r^2}{\omega_{z}^2}\right)\exp{\left(-\frac{r^2}{\omega_z^2}-j\tilde{\nu}z\right)}\times
        \\ 
        &\exp{\left(-j\tilde{\nu}\frac{r^2}{2R_z}+j\left(2p_\mathrm{I}+\left | l_\mathrm{I} \right |+1\right)\zeta\left(z\right)-jl_\mathrm{I}\theta\right)},
    \end{split}
    \label{Eq.LGModes}
\end{equation}
where $\tilde{\nu}$ = $2\pi/\lambda$ stands as the wavenumber, $p_\mathrm{I}$ and $l_\mathrm{I}$ represent the radial index and azimuth index of the $i$th transmitted mode, respectively. The term $\omega_{z}$ indicates the radius of the beam spot, which can be calculated as $\omega_{z} = \omega_{0} \sqrt{(1+2p_{\mathrm{I}}+|l_{\mathrm{I}}|)( 1+z/z_{R})}$, with $\omega_{0} $ and $z_{R}$ being the waist radius and Rayleigh range, respectively. $R_{z}$ and $\zeta \left( z\right)$ define the curvature radius of the beam's wavefront and the Gouy phase at a distance of $z$, respectively.

Practical transceiver misalignment can result in a dual set of impairments: pointing error and AOA fluctuations. The fluctuations in the transmitter's orientation angle lead to the displacement between the optical fields and the receiver's aperture, consequently causing pointing error. Fig. \ref{Displacement1} depicts the pointing error induced by the transmitter vibration, highlighting a displacement $\vec{d}$ between the beam center and the aperture center. Without loss of generality, we assume $\vec{d}$ aligns with the $x$-axis. In this context, the radial vector $\Vec{r}$ in the receiver plane essentially indicates the position of $\Vec{r'}$ relative to the transmitter coordinate system. After the coordinate transformation, the $\Vec{r'}$ is with a magnitude of $g(r,\theta) = \sqrt{r^2+d^2-2rd\cos{\theta}}$ and the azimuth angle of $\vartheta$. The relationship between $\vartheta$ and $\theta$ is given by
\begin{equation}
    \vartheta=
    \begin{cases}
        \pi-\arccos{\left(\frac{d-r\cos{\theta}}{\sqrt{r^2+d^2-2rd\cos{\theta}}}\right)}, & {\theta \in [0,\pi]},
        \\
         \pi+\arccos{\left(\frac{d-r\cos{\theta}}{\sqrt{r^2+d^2-2rd\cos{\theta}}}\right)}, & {\theta \in [\pi,2\pi]}.
    \end{cases}
\end{equation}
While the existing work has emphasized the misaligned field impaired by the lantern displacement \cite{SlantLGBeams,FMFCouplingDis}, they generally neglect the potential impacts on the azimuthal dependence which is fundamental to the functioning of SMM systems. Therefore, after accounting for the transformation of both radial distance $|\Vec{r'}|$ and the azimuth angle $\vartheta$ induced by the pointing error $\Vec{d}$, the field distribution of the incident LG mode over the receiver lens becomes
\begin{align}
&
{E^{(i)}_{\mathrm{I}}}\left(\Vec{r'},z\right) = {E^{(i)}_{\mathrm{I,A}}}\left(\Vec{r},z,\Vec{d}\right) = \mathrm{LG}_{p_\mathrm{I},l_\mathrm{I}}\left(r,\theta,z,d\right) = \nonumber
\\&\frac{B^\mathrm{I}_{pl}}{\omega_{z}}\left(\sqrt{2}\ \frac{g(r,\theta)}{\omega_{z}}\right)^{\left | l_\mathrm{I} \right |}L^{\left | l_\mathrm{I} \right |}_{p_\mathrm{I}}\left(2\ \frac{g^2(r,\theta)}
{\omega_{z}^2}\right)\exp{\left(-\frac{g^2(r,\theta)}{\omega_z^2}\right)}\times \nonumber
\\
&\exp{\left(-j\tilde{\nu}\frac{g^2(r,\theta)}{2R_z}-j\tilde{\nu}z+j\left(2p_\mathrm{I}+\left | l_\mathrm{I} \right |+1\right)\zeta\left(z\right)-j\left | l_\mathrm{I} \right |\vartheta\right)}.
\label{Eq.LGModesDis}
\end{align}

\begin{figure}[t]
   \centering
    \includegraphics[scale=0.6]{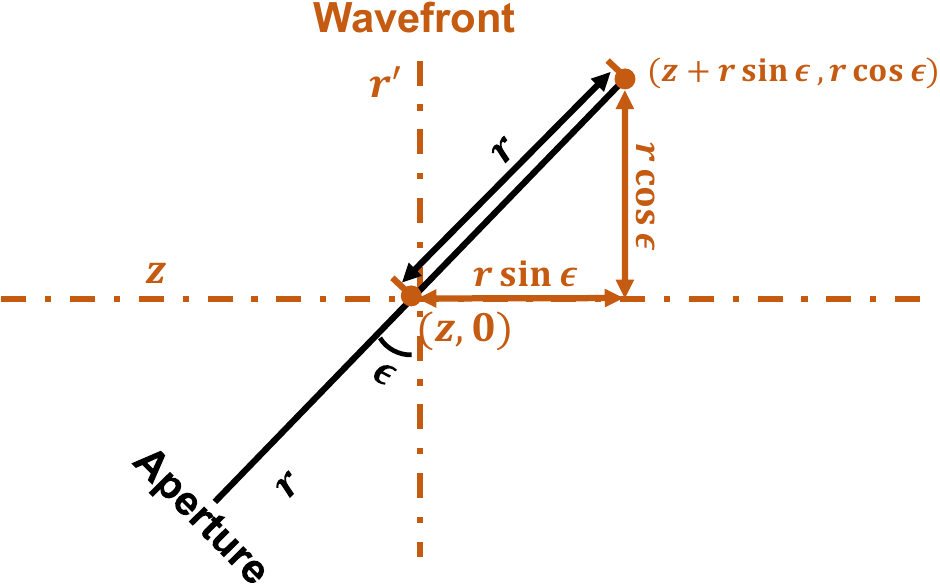}
    \vspace{5mm}
    \caption{Illustration of the receiver vibration-induced AOA fluctuation, showing the cross-section of the wavefront intersecting the aperture at an angle $\varepsilon$ at distance $z$.}
    \label{AOA}
\end{figure}
On the other hand, receiver vibrations can give rise to fluctuations in the spatial phase of the near-field spherical beam across the aperture, commonly referred to as the AOA fluctuations \cite{Nearfield,AOA}. For a receiver aperture which is significantly smaller than the beam's coherence spot, the wavefront is essentially a plane over the aperture \cite{AOA}. As described in \eqref{Eq.LGModes}, when the incident wavefront is perfectly parallel with the receiver aperture, i.e., signifying no AOA deviations, the spatial phase of the $i$th beam across the aperture can be expressed as
\begin{equation}
    \varphi_i\left(r',z\right) = \varphi_i\left(r,z\right) = \frac{r^2}{2R_z}+\tilde{\nu}z-\left(2p_\mathrm{I}+\left | l_\mathrm{I} \right |+1\right)\zeta\left(z\right)+ l_\mathrm{I}\theta.
\end{equation}
However, given that the incident wavefront has an angle of $\varepsilon$ with the aperture plane due to the AOA fluctuations, as depicted in Fig. \ref{AOA}, the phase distribution across the aperture becomes
\begin{align}
&\varphi_i\left(r',z\right) =\varphi_i\left(r,z,\varepsilon\right) = \frac{\left(r\cos{\varepsilon}\right)^2}{2R_{z+r\sin{\varepsilon}}}+ \nonumber
\\
&\tilde{\nu}(z+r\sin{\varepsilon})-\left(2p_\mathrm{I}+\left | l_\mathrm{I} \right |+1\right)\zeta\left(z+r\sin{\varepsilon}\right)+l_\mathrm{I}\theta.
\label{Eq.PhaseDisWoSimplified}
\end{align}
Assuming the typically small deviation angle $\varepsilon$, as observed in measurements from real-world vibrations \cite{Datacentervibration}, the small angle approximation is justified. This permits the assumptions that $\cos{\varepsilon} \approx 1$ and $\sin{\varepsilon} \approx \varepsilon$. Additionally, taking into account the slow variation of both the wavefront curvature radius, $R_z$, and the Gouy phase, $\zeta(z)$, with respect to the distance $z$ \cite{PhotonicsBook}, \eqref{Eq.PhaseDisWoSimplified} can be simplified to
\begin{equation}
    \varphi_i\left(r,z,\varepsilon\right) = \frac{r^2}{2R_{z}}+\tilde{\nu}z+\tilde{\nu}r\varepsilon-\left(2p_\mathrm{I}+\left | l_\mathrm{I} \right |+1\right)\zeta\left(z\right)+ l_\mathrm{I}\theta.
    \label{Eq.PhaseDisSimplified}
\end{equation}
Subsequently, by substituting \eqref{Eq.PhaseDisSimplified} into \eqref{Eq.LGModesDis}, the incident field distribution across the aperture, in the presence of both pointing error denoted by $d$, and AOA fluctuations represented by $\varepsilon$, can be articulated as
\begin{align}
     &{E^{(i)}_{\mathrm{I}}}\left(\Vec{r'},z\right)={E^{(i)}_{\mathrm{I,A}}}\left(\Vec{r},z,\Vec{d},\varepsilon \right)=\mathrm{LG}_{p_\mathrm{I},l_\mathrm{I}}\left(r,\theta,z,d,\varepsilon\right) = \nonumber
     \\
     &\frac{B^\mathrm{I}_{pl}}{\omega_{z}}\left(\sqrt{2}\ \frac{g(r,\theta)}{\omega_{z}}\right)^{\left | l_\mathrm{I} \right |}L^{\left | l_\mathrm{I} \right |}_{p_\mathrm{I}}\left(2\ \frac{g^2(r,\theta)}
     {\omega_{z}^2}\right)\times\nonumber
     \\
    &\exp{\left(-\frac{g^2(r,\theta)}{\omega_z^2}-j\tilde{\nu}\frac{g^2(r,\theta)}{2R_z}-j\tilde{\nu}\left(z+\varepsilon g(r,\theta)\right)\right)}\times\nonumber
    \\
    &\exp{\left(j\left(\left(2p_\mathrm{I}+\left | l_\mathrm{I} \right |+1\right)\zeta\left(z\right)-l_\mathrm{I} \vartheta\right)\right)}.
    \label{Eq.LGmodesMis}
\end{align}

By utilizing the derived incident field across the aperture, the fiber-coupled power across different modes in the presence of misalignment can be calculated. Specifically, by substituting \eqref{Eq.EstimatedFiberModes} and \eqref{Eq.LGmodesMis} into \eqref{Eq.CouplingPower2}, and following a series of algebraic manipulations, the final coupling power between the $i$th transmitted $\mathrm{LG}_{p_\mathrm{I},l_\mathrm{I}}$ mode and the $k$th fiber mode ($\mathrm{LP}_{l_\mathrm{F},p_{\mathrm{F}+1}}$), incorporating the impact of transceiver vibrations, can be expressed as
\begin{align}
    &\left|h_{ik}\right|^2= {P^{(ik)}_{\,\mathrm{F}}}  = \frac{\left({B^\mathrm{F}_{pl}}{B^\mathrm{I}_{pl}}\alpha\right)^2}{2}\exp{\left(-\frac{2d^2}{\omega^2(z)}\right)}\left(2\beta^2\right)^{\left | l_\mathrm{F} \right |+1}\times\nonumber
    \\
    &\left |  \int\limits_{0}^{1} \int\limits_{0}^{2\pi}  \left(\frac{\sqrt{2} f(x,\theta)}{\omega_z}\right)^{\left | l_\mathrm{I}\right |} L_{P_\mathrm{I}}^{\left | l_\mathrm{I}\right |}\left( \frac{2 f^2(x,\theta)}{\omega_z^2}\right) L_{P_\mathrm{F}}^{\left | l_\mathrm{F}\right |}\left(2\beta^2x^2\right)\right.\times  \nonumber \\
    &\left.\exp{\left(-j\tilde{\nu}\frac{xDd\cos{\theta}}{2R_z}+j\tilde{\nu}\varepsilon f\left(x,\theta\right) -\gamma x^2 \right)} x^{\left | l_\mathrm{F}\right |+1}\right.\times  \nonumber \\
    &\left.\exp{\left(\frac{xDd\cos{\theta}}{\omega^2_z}+jl_\mathrm{I}\vartheta-jl_\mathrm{F} \theta \right)} \mathrm{d}x \mathrm{d}\theta\right |^2,
    \label{Eq.CoupledPowerFinal}
    \end{align}
where the function $f(x,\theta) = (\frac{D^2}{4}x^2+d^2-xDd\cos{\theta})^{\frac{1}{2}}$, and $\gamma$ is defined as $\gamma = \alpha^2+\beta^2-j\frac{\tilde{\nu}D^2}{8R_Z}$, with $\alpha = \frac{D}{2\omega_Z}$ representing the ratio of the aperture radius to the incident beam radius. Moreover, $\beta = \frac{D}{2\omega}$ signifies the ratio of the aperture radius to the back-propagated waist radius, which is commonly referred to as the coupling geometry parameter.

\subsection{Coupling Efficiency in the Presence of Misalignment}
The coupling efficiency denotes the ratio between the coupled power into the fiber and the collected power by the aperture, which can be expressed as
\begin{equation}
     \eta_{ik}  =  \frac{ {P^{(ik)}_{\,\mathrm{F}}}}{ {P^{(i)}_{\,\mathrm{A}}}} = \frac{| h_{ik} |^2 }{ \int_{A} | {E^{(i)}_{\mathrm{I,A}}} \left ( \vec{r}  \right )  |^2 \mathrm{d} \vec{r} },
    \label{Eq.CouplingEfficiencyIntial}
\end{equation}
where the term ${P^{(i)}_{\,\mathrm{A}}}$ denotes the collected power by the receive aperture from $i$th incident mode. Substituting   \eqref{Eq.LGmodesMis} and \eqref{Eq.CoupledPowerFinal} into \eqref{Eq.CouplingEfficiencyIntial} and following a series of algebraic manipulations, the final coupling efficiency expression for the $k$th fiber mode with the $i$th incident mode can be expressed as
\begin{align}
    &\eta_{ik} = {B^{\mathrm{F}}_{pl}}^2\left(2\beta^2\right)^{\left | l_{\mathrm{F}} \right |+1}\left | \int\limits_{0}^{1} \int\limits_{0}^{2\pi}  
    \exp{\left( -\gamma x^2 + j\tilde{\nu}\varepsilon f\left(x,\theta\right)\right)} \right.\times  \nonumber \\
    &\left.\exp{\left(xDd\cos{\theta}\left(\frac{2R_z-j\tilde{\nu}\omega^2_z}{2\omega^2_z R_z}\right)+j\left(l_{\mathrm{I}}\vartheta-l_{\mathrm{F}} \theta\right)\right)}x^{\left | l_{\mathrm{F}}\right |+1}\right.\times\nonumber
    \\
    &\left.\left(\frac{\sqrt{2} f(x,\theta)}{\omega_z}\right)^{\left | l_{\mathrm{I}}\right |} L_{P_{\mathrm{I}}}^{\left | l_{\mathrm{I}}\right |}\left( \frac{2 f^2(x,\theta)}{\omega_z^2}\right)   L_{P_{\mathrm{F}}}^{\left | l_{\mathrm{F}}\right |}\left(2\beta^2x^2\right)\mathrm{d}x \mathrm{d}\theta\right |^2 
    \times\nonumber
    \\
    &\left(2\int\limits_{0}^{1} \int\limits_{0}^{2\pi}\left(\frac{2 f^2(x,\theta)}{\omega_z}\right)^{\left | l_{\mathrm{I}}\right |}\left(L_{P_{\mathrm{I}}}^{\left | l_{\mathrm{I}}\right |}\left( \frac{2 f^2(x,\theta)}{\omega_z^2}\right)\right)^2 \right.\times\nonumber
    \\
    &\left.\exp{\left(2\left(\frac{xDd\cos{\theta}}{\omega^2_z}-\alpha^2x^2\right)\right)}x\mathrm{d}x\mathrm{d}\theta \right)^{-1}.
    \label{Eq.CouplingEfficiency}
    \end{align}
Under the far-field assumption, considering that the input optical intensity is independent of $\Vec{r}$ across the receiver aperture, the coupling efficiency of the incident Gaussian beam with SMF in the absence of link misalignment is reduced to
\begin{equation}
\eta=8 \beta^2\left|\int_0^1 x \exp \left(-\beta^2 x^2\right) \mathrm{d} x\right|^2,
\end{equation}
which aligns with the literature with negligible speckle size as depicted in \cite{SMFCouplingEfficiency1, SMFCouplingEfficiency2}.

Assuming that both the vertical and horizontal AOA deviations, denoted by $\varepsilon_v$ and $\varepsilon_h$ respectively, follow independent and identical zero-mean Gaussian distribution, the combined AOA deviation under the small angle approximation is $ \varepsilon = \sqrt{\varepsilon_v^2+\varepsilon_h^2}$. This leads to a Rayleigh distribution characterized by a scale parameter of $\sigma_{\varepsilon}$ with the probability density function (PDF)
\begin{equation}
p_\varepsilon(\varepsilon)=\frac{\varepsilon}{\sigma_{\varepsilon}^2} \exp \left(-\frac{\varepsilon^2}{2 \sigma_{\varepsilon}^2}\right).
\end{equation}
Similarly, by considering identical distributions for vertical and horizontal transmitter orientation angles, denoted as $\epsilon _v$ and $\epsilon _h$ respectively, the resultant vertical and horizontal displacements can be modeled as $d_v$ = $z \epsilon _v$ and $d_h$ = $z \epsilon _h$. Consequently, the aggregate displacement $d = z\sqrt{\epsilon _v^2+\epsilon _h^2}$ is also subject to a Rayleigh distribution with a scale parameter $\sigma_{d} = z\sigma_{\epsilon}$, and can be represented as $p_d(d)$. Therefore, the expected coupling efficiency $\left \langle \eta_{ik} \right \rangle$ can be calculated by integrating (\ref{Eq.CouplingEfficiency}) over the effects of both pointing error and AOA deviations, which is given by
 \begin{equation}
    \left \langle \eta_{ik} \right \rangle = 
    \int\limits_{-\infty}^{\infty}\int\limits_{-\infty}^{\infty}\eta_{ik}(d,\varepsilon)p_d(d)p_\varepsilon(\varepsilon)\mathrm{d}\varepsilon\mathrm{d}d,
\end{equation}
where the angle brackets represent the ensemble-average operator.

\subsection{Achievable Data Rate}
Building upon the channel coefficients derived in \eqref{Eq.CoupledPowerFinal}, the estimated complex channel matrix for the considered FMF-coupled SMM system can be expressed as
\begin{equation}
    \mathbf{\Tilde{H}} = \left[\begin{array}{ccc}
\left|h_{11}\right| & \ldots & \left|h_{N 1}\right|e^{j\Delta \varphi^1_{N-1}} \\
\vdots & \ddots & \vdots \\
\left|h_{1 N}\right|e^{j\Delta \varphi^N_{1-N}} & \ldots & \left|h_{N N}\right|
\end{array}\right],
\label{Eq.EstChannelMatrix}
\end{equation}
where $\Delta \varphi^k_{n-i}$ denotes the phase difference between $h_{ik}$ and $h_{nk}$.  In \cite{IMDDDataRate}, it was shown that if ${s}_k$ obeys the Rayleigh distribution and ${s}_k^2$ obeys the exponential distribution, a capacity lower bound for optical IM/DD channel can be derived. Using this result, an achievable data rate for $k$th subchannel  can be expressed as  
\begin{equation}
    r_k = \frac{1}{2}B\log{\left(1+\frac{\varsigma_k e}{2\pi}\right)},
    \label{Eq.SubchannelDataRate}
\end{equation}
where $B$ stands for the bandwidth, and $\varsigma_k$ indicates the signal-to-interference-plus-noise ratio (SINR) for channel $k$. In real-world demonstrations, a total power constraint needs to be considered due to issues such as eye-safety requirement, which can be expressed as $\sum \xi_k \leq \xi_\mathrm{t}$ \cite{Eyesafety}, where $\xi_k = \langle {s}_k^2 \rangle$ denotes the average optical power for $k$th channel and $\xi_\mathrm{t}$ is the total power limit. Considering that significant interference is introduced by the channel misalignment, according to the non-linear channel transformation for the mutually coherent system as depicted in \eqref{Eq.ChannelTrans}, the average received current for $k$th PD, $\langle Y_k \rangle$, can be denoted as

\begin{align}
    &\langle Y_k \rangle = R \left\langle \left| \sum_{i=1}^N h_{ik}{s_i}\right|^2\right\rangle = R\sum_{i=1}^N\sum_{n=1}^N h_{ik} h_{nk}^* \left\langle{s_i}{s_n}\right\rangle\nonumber
    \\
    & = R\left( \left|h_{kk}\right|^2\xi_k+\sum_{i=1\atop i\neq k}^{N}\left|h_{ik}\right|^2\xi_i+\sum_{i=1}^N\sum_{n=1\atop n\neq i}^{N}\frac{\pi}{4}h_{ik}h_{nk}^* \sqrt{\xi_i \xi_k}\right),
\end{align}
where the first element illustrates the desired signal emitted by the $k$th transmitter, and the second and third elements indicate the interference and beat noise, respectively. Note that since ${s}_k^2$ is assumed to be exponentially distributed, we have $\langle {s}_k^2 \rangle = \xi_k$ and $\langle {s}_k \rangle = \sqrt{\xi_k \pi}/2$. With the uniform power allocation among total $N$ transmitters, i.e., $\xi_1 = \dots = \xi_N = \frac{\xi_\mathrm{t}}{N}$, after using $X_k$ to collectively denote the interference and beat noise, the $\varsigma_k$ is given by
\begin{equation}
    \varsigma_k = \frac{R^2\xi_{\mathrm{t}}^2\left|h_{kk}\right|^4}{N^2\left(\sigma^2_n+ X_k^2\right)} = \frac{R_fR^2\xi_{\mathrm{t}}^2\left|h_{kk}\right|^4}{N^2\left(4k_bTF_nB+R_fX_k^2\right)},
    \label{Eq.SINR}
\end{equation}
with $k_b$ being the Boltzmann constant, $R_f$ being the feedback resistor for the receiver transimpedance amplifier (TIA), $F_n$ being the noise figure and $T$ being the temperature in Kelvin. 

Following the substitution of \eqref{Eq.SINR} into \eqref{Eq.SubchannelDataRate}, and taking into account the sum of achievable data rates across all subchannels as $C = \sum_k r_k$, the aggregated channel capacity in the presence of interference can be described as
\begin{equation}
    C_{\mathrm{I}} =  \sum_{k=1}^N\frac{1}{2}B\log{\left(1+\frac{e R_f R^2\xi_t^2\left|h_{kk}\right|^4 }{2\pi N^2\left(4k_bTF_nB\pi+R_fX_k^2\right)}\right)}.
    \label{Eq.CapacitywoZFBF}
\end{equation}

As depicted in \eqref{Eq.CapacitywoZFBF}, the presence of interference and beat noise precipitates significant performance degradation, underscoring the imperative need for a robust interference mitigation strategy to ensure optimal signal integrity and system efficiency. The investigation into the non-linear transformation has demonstrated that traditional postcoding techniques are inadequate for mitigating interference arising from subchannels \cite{MutuallyCohernet, ShenjieSMMZFBF,Phaseretravel1}. Nonetheless, given the availability of channel state information at the transmitter end, ZFBF can be used as a viable solution to achieve crosstalk-free parallel transmission. This is particularly relevant under high-speed transmission, which suggests the channel matrix $\mathbf{H}$ is assumed to be quasi-static during the transmission of a symbol block. Observing from \eqref{Eq.EstChannelMatrix}, the relationship between the channel matrix $\mathbf{H}$ and $\mathbf{\Tilde{H}}$ can be revealed as $\mathbf{H} = \mathbf{D}\mathbf{\Tilde{H}}$, with $\mathbf{D}$ being the diagonal matrix with only phase elements. Therefore, if we choose the inverse of the estimated channel matrix as the precoder, the received electrical current matrix $\mathbf{Y}$ is given by
\begin{equation}
    \mathbf{Y} = R\left| \mathbf{H}\mathbf{\Tilde{H}}^{-1}\mathbf{\Tilde{s}} \right|^2+\mathbf{Z} =  R\left| \mathbf{\Tilde{s}} \right|^2+\mathbf{Z},
\end{equation}
where $\mathbf{\Tilde{s}} = [\Tilde{s_1}, \Tilde{s_2}, ..., \Tilde{s_N}]$ indicates the information bearing signal. Assuming that $\Tilde{s}_k$ also obeys the Rayleigh distribution, after the crosstalk compensation introduced by the ZFBF, the signal-to-noise ratio (SNR) for $k$th subchannel is given by
\begin{equation}
    \varsigma'_k = \frac{\left(R\Tilde{\xi_k}\right)^2}{\sigma^2_n} = \frac{R_fR^2\Tilde{\xi_k}^2}{4k_bTF_nB}.
    \label{Eq.SNR}
\end{equation}
With the non-linear power allocation induced by the ZFBF, the average total transmitted power, $\langle||\mathbf{\Tilde{s}}||^2\rangle$, which satisfies $\langle||\mathbf{\Tilde{s}}||^2\rangle \leq \xi_\mathrm{t}$, can be derived as
\begin{align}
    &\langle||\mathbf{\Tilde{s}}||^2 \rangle= \left\langle \sum_{k=1}^N\left| \sum_{i=1}^N I_{ik} \Tilde{s_i}\right|^2\right\rangle = \sum_{k=1}^N\sum_{i=1}^N\sum_{n=1}^N I_{ik} I_{nk}^* \left\langle\Tilde{s_i} \Tilde{s_n}\right\rangle\nonumber 
    \\ &=\sum_{k=1}^N\sum_{i=1}^N\left(\left|I_{ik}\right|^2\Tilde{\xi}_i+\frac{\pi}{4}\sum_{n=1 \atop n\neq i}^{N}{I_{ik} I_{nk}^*\sqrt{\Tilde{\xi_i} \Tilde{\xi_n}}}\right),
    \label{Eq.ZFBFPower}
\end{align}
where $I_{ik}$ stands for the elements in the inverse of estimated channel matrix $\mathbf{\Tilde{H}}$. Therefore, considering the uniform power allocation among the subchannels, i.e., $\Tilde{\xi_1} = \dots = \Tilde{\xi_N} = \Tilde{\xi_\mathrm{un}}$, and that capacity lower bound is achieved at the boundary, i.e., $\langle||\mathbf{\Tilde{s}}||^2\rangle = \xi_\mathrm{t}$, we can obtain $\Tilde{\xi_\mathrm{un}}$ from \eqref{Eq.ZFBFPower} as
\begin{equation}
    \Tilde{\xi_\mathrm{un}}= \frac{\xi_\mathrm{t}}{\sum_{k=1}^N\sum_{i=1}^N\left(\left|I_{ik}\right|^2+\frac{\pi}{4}\sum_{n=1, n\neq i}^{N}{I_{ik} I_{nk}^*}\right)}.
    \label{Eq.ZFBFUniPowerAllo}
\end{equation}
Consequently, by following the substitution of \eqref{Eq.ZFBFUniPowerAllo} and \eqref{Eq.SNR} into \eqref{Eq.SubchannelDataRate}, the total channel capacity with the ZFBF can be revealed as
\begin{equation}
    C_{\mathrm{ZF}} =  \frac{N}{2}B\log{\left(1+\frac{e R_f R^2\Tilde{\xi_\mathrm{un}}^2}{8k_bTF_nB\pi}\right)}.
    \label{Eq.CapacitywZFBF}
\end{equation}

\section{Simulation results and discussion}
\label{SimulationResults}
\begin{figure*}[t]
   \centering
   \subfloat[]{
    \includegraphics[scale=0.45]{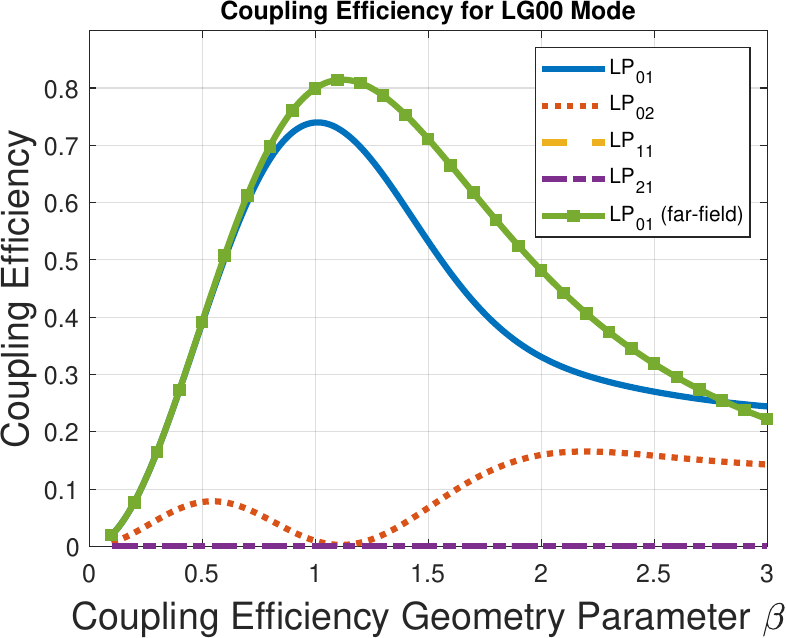}
    \label{LG00NoMisalignment}}
       \subfloat[]{
    \includegraphics[scale=0.45]{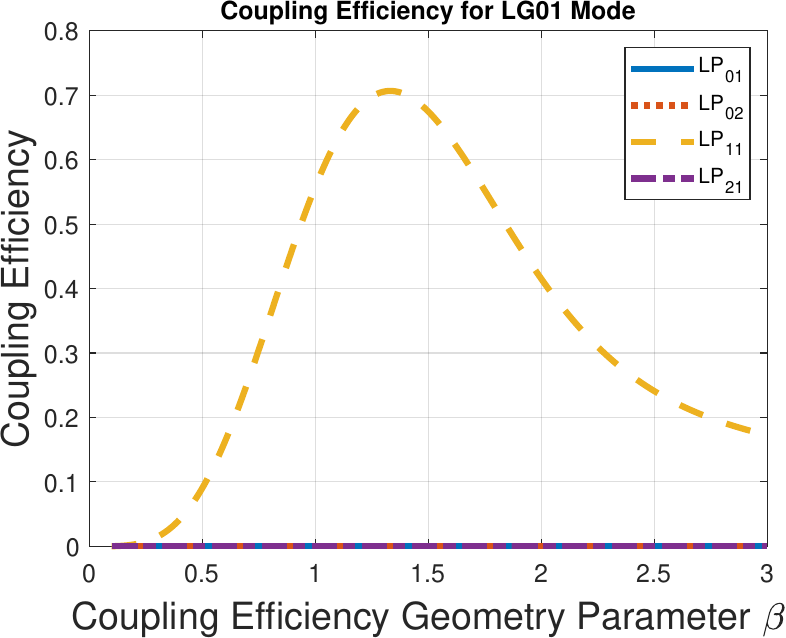}
    \label{LG01NoMisalignment}}
       \subfloat[]{
    \includegraphics[scale=0.45]{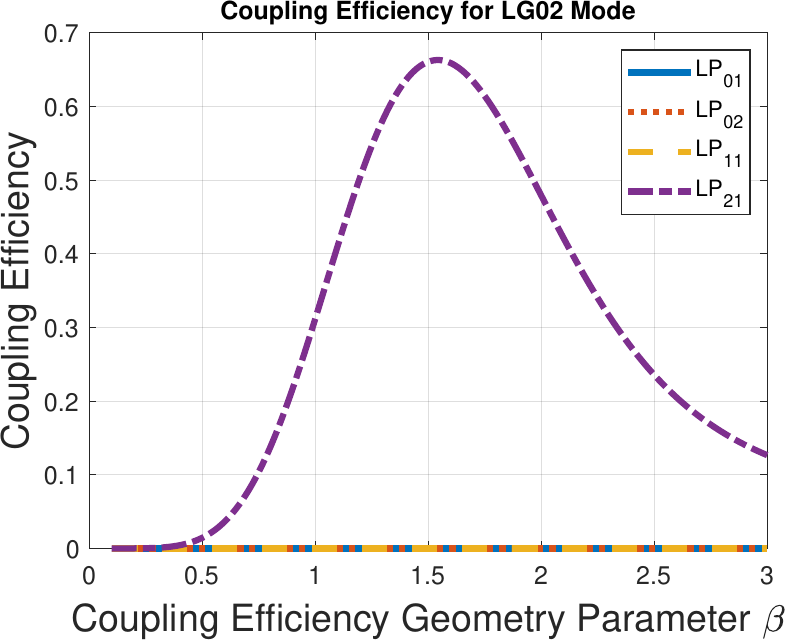}
    \label{LG02NoMisalignment}}
    \\
    \subfloat[]{
    \includegraphics[scale=0.45]{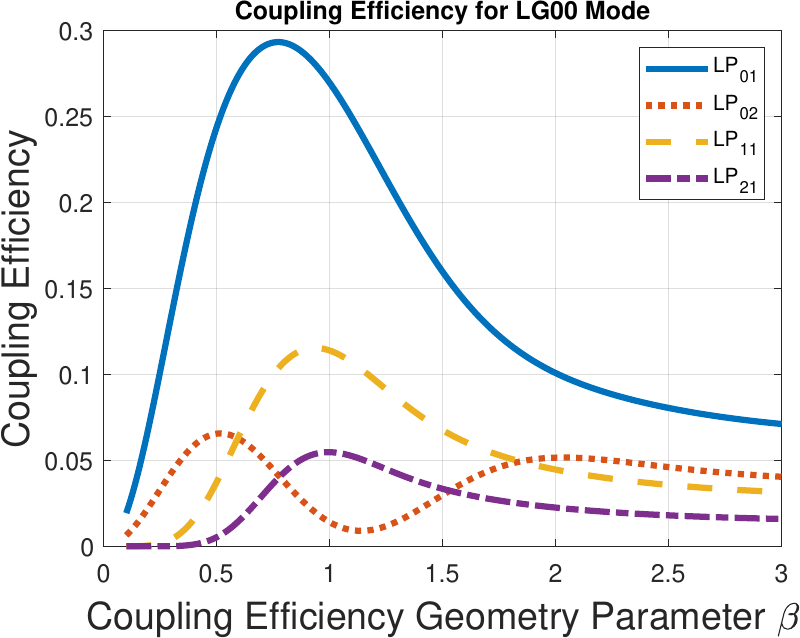}
    \label{LG00FixedMisalignment}}
       \subfloat[]{
    \includegraphics[scale=0.45]{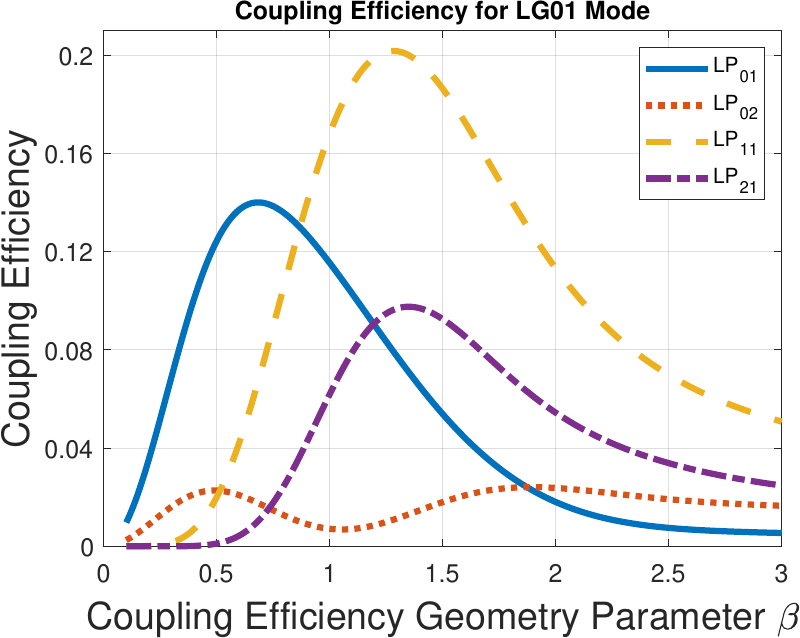}
    \label{LG01FixedMisalignment}}
       \subfloat[]{
    \includegraphics[scale=0.45]{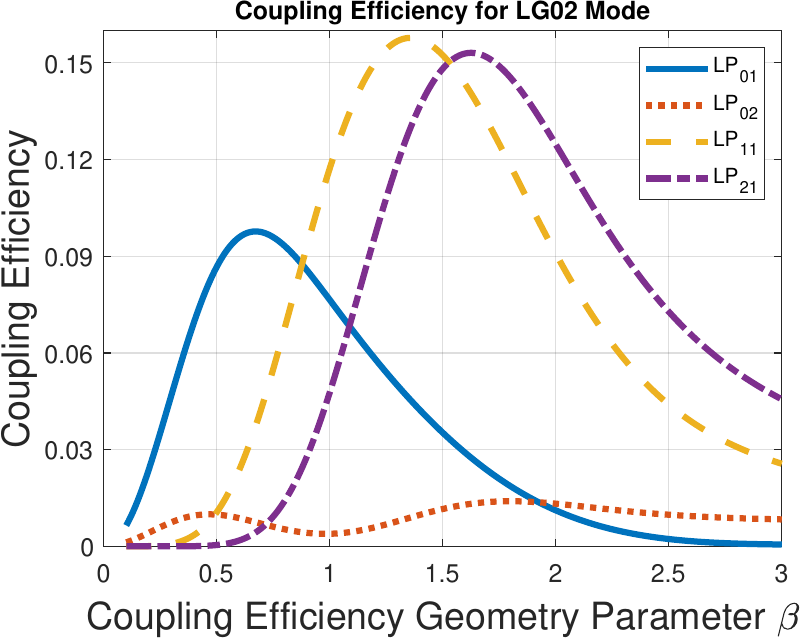}
    \label{LG02FixedMisalignment}}
    \caption{The coupling efficiency for a six-mode FMF as the function of coupling parameter $\beta$ in the absence of misalignment for an incident (a): $\mathrm{LG}_{00}$ mode, (b): $\mathrm{LG}_{01}$ mode, (c): $\mathrm{LG}_{02}$ mode; and in the presence of misalignment for an incident (d): $\mathrm{LG}_{00}$ mode, (e): $\mathrm{LG}_{01}$ mode, (f): $\mathrm{LG}_{02}$ mode.}
    \label{SimCouplingEfficiency}
\end{figure*}
In this section, we will evaluate the performance of the SMM-enabled FMF coupling OWC system in the presence of misalignment, including both coupling efficiency and communication performance. In the simulation, the transmitted beam is considered with a wavelength of $1550$ $\mathrm{nm}$ and a waist radius of $\omega_o = 800$ $\mathrm{\mu m}$. A link distance of $10$ $\mathrm{m}$ is assumed to emulate a typical indoor laser-based OWC link such as in data centers \cite{Datacentervibration}. The total transmission power constraint $\xi_t$ is set as 10 dBm \cite{Eyesafety}, unless otherwise indicated. The other employed parameters include a feedback resistor value of $R_f = 500$ $\Omega$, a noise figure of $F_n = 5$ dB, a stable ambient temperature of $T = 300^\circ$ K, and a receiver responsivity of $R = 0.7$ A/W \cite{FOVDataRateTradeoff}.
\subsection{Coupling Efficiency}
We first assess the coupling efficiency as the function of $\beta$ considering different incident OAM modes ($\mathrm{LG}_{00}$, $\mathrm{LG}_{01}$, and $\mathrm{LG}_{02}$) without link misalignment, where $\beta$ is the coupling geometry parameter which denotes the ratio of the aperture radius to the back-propagated waist radius. This assessment is conducted in a six-mode FMF supporting $\mathrm{LP}_{01}$, $\mathrm{LP}_{02}$, $\mathrm{LP}_{11}$, and $\mathrm{LP}_{21}$ modes. As illustrated in Fig. \ref{LG00NoMisalignment} Fig. \ref{LG01NoMisalignment} and Fig. \ref{LG02NoMisalignment}, the coupled power between the fiber mode $\mathrm{LP}_{l_\mathrm{F},p_\mathrm{F}+1}$ (approximated by $\mathrm{LG}_{p_\mathrm{F},l_\mathrm{F}}$), and an incident beam, $\mathrm{LG}_{p_\mathrm{I},l_\mathrm{I}}$,  with different azimuth order (i.e., $l_{\mathrm{I}} \neq l_{\mathrm{F}}$) is found to be zero. This result underscores that the orthogonality between modes with distinct azimuthal phase dependencies has been well maintained. It is also observed that, in Fig. \ref{LG00NoMisalignment}, the coupling efficiency of $\mathrm{LP}_{01}$ attains its maximal value of 0.74 when $\beta$ is at 1.01. Under the far-field assumption, the coupling efficiency can increase to 0.81 when $\beta$ is 1.12, a result that aligns with the findings in the existing literature \cite{SMFCouplingEfficiency1,SMFCouplingEfficiency2}. On the other hand, $\mathrm{LP}_{02}$ mode is less excited compared with $\mathrm{LP}_{01}$ mode due to the field vectors reverse \cite{FMFCouplingDis}, with a maximum coupling efficiency of 0.166 when $\beta$ is 2.2. A notable fluctuation in coupling efficiency for the mode $\mathrm{LP}_{02}$ is observed with increasing $\beta$, which is linked to the intensity profile of $\mathrm{LP}_{0p}$ modes with a central spot surrounded by $p-1$ concentric rings when $p \geq 2$ \cite{PhotonicsBook}. Given that $\omega$ is held constant in our simulation, an increase in the coupling parameter $\beta$ leads to a larger aperture radius. This enlargement of the aperture radius affects the spatial separation between the central spot and the surrounding ring in the $\mathrm{LP}_{02}$ mode, leading to a fluctuated coupling performance. As illustrated in Fig. \ref{LG01NoMisalignment}, the calculated optimal $\beta$ stands at 1.33, which is associated with a maximum coupling efficiency of 0.71 for the incident $\mathrm{LG}_{01}$ mode. In the case of the incident $\mathrm{LG}_{02}$ beam in Fig. \ref{LG02NoMisalignment}, a $\beta$ setting of 0.66 is found to facilitate the highest coupling efficiency which is recorded at 1.54.

Subsequently, we investigate the coupling performance under conditions of both pointing error and AOA fluctuations, focusing on the previously mentioned transmitted modes set and the FMF. Notably, we adopt a parameter of 0.125 $\mathrm{mrad}$ for the Rayleigh distribution ($\sigma_{\epsilon}$ and $\sigma_{\varepsilon}$), aligning with real-world scenarios \cite{Datacentervibration}. As depicted in Fig. \ref{LG00FixedMisalignment} Fig. \ref{LG01FixedMisalignment} and Fig. \ref{LG02FixedMisalignment}, the deleterious effects of transceiver vibrations significantly degrade the coupling efficiency, and the performance degradation escalates with the increase of the transmitted mode orders. For example, with the incident $\mathrm{LG}_{00}$ beam, the ideal $\beta$ has shifted to 0.77 to attain the highest coupling efficiency for the $\mathrm{LP}_{01}$, which is 0.29 and with a 61\% decrease. As for the incident $\mathrm{LG}_{01}$ mode, the maximum coupling efficiency has been diminished by 72\%, to a value of 0.20 with the optimal $\beta$ being 1.28. For the incident $\mathrm{LG}_{02}$ beam, the optimal $\beta$ is 1.63 now, leading to the maximum coupling efficiency of only 0.15 after a notable 77\% drop. Moreover, in the presence of the misalignment, the orthogonality among modes with different azimuthal dependence can not be conserved, aligning with the power leakage observed in previous research\cite{LGBEAMDis,OAMDis}. This phenomenon occurs because, in the absence of misalignment, the transmitted mode is perfectly out of phase with half of the modes with a differing azimuthal order. Such a perfect phase mismatch leads to a cancellation effect, which prevents the excitation of the remaining modes in the set. Considering the link misalignment introduces a phase change to the incident beam before being coupled into the fiber, and the induced imperfect phase mismatch results in the mode excitation \cite{FMFCouplingDis}. In addition, it is observed that the power of the transmitted mode is more likely to be coupled into the supported mode with the nearest mode order under the considered misalignment. Specifically, Fig. \ref{LG00FixedMisalignment} depicts that with the $\beta$ of 0.77, after being normalized by the aperture collected power, 11$\%$ of the power is leaked into the $\mathrm{LP}_{11}$ mode and 5$\%$ power is coupled into $\mathrm{LP}_{21}$ mode. Similar trends can also be observed in Fig. \ref{LG01FixedMisalignment} and Fig. \ref{LG02FixedMisalignment}, namely,  10$\%$ of the power is leaked into the $\mathrm{LP}_{01}$ mode and 8$\%$ power is coupled into $\mathrm{LP}_{21}$ mode with the incident $\mathrm{LG}_{01}$ beam; 15$\%$ of the power is leaked into the $\mathrm{LP}_{01}$ mode and 3$\%$ power is coupled into $\mathrm{LP}_{21}$ mode with the incident $\mathrm{LG}_{02}$ beam. Consequently, it is deduced that transceiver vibrations not only substantially diminish coupling efficiency but also severely disrupt the orthogonality among distinct spatial modes, and the severity of the deleterious impact is amplified with higher-order LG beams.

\subsection{SMM System Communication Performance}
\begin{figure}[t]
    \centering
    \includegraphics[scale=0.63]{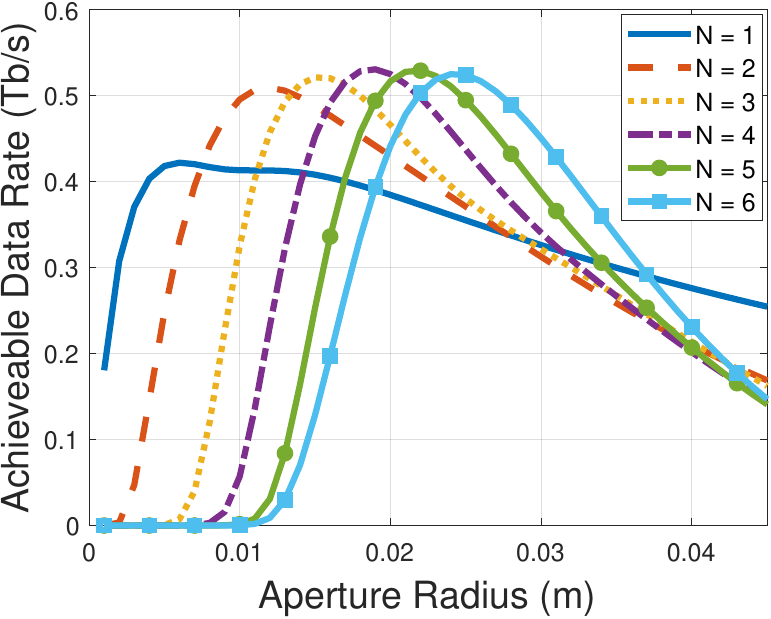}
    \caption{ The achievable data rate as the function of aperture radius under different $N$ where the focal length $f$ equals to the aperture diameter $D$.}
    \label{DataDiffR}
\end{figure}

In this section, the communication performance of the ZFBF-based SMM system will be evaluated. It should be noted that the achievable data rate for each channel within the selected transmitted mode set can be quantitatively determined by applying \eqref{Eq.CapacitywoZFBF} and \eqref{Eq.CapacitywZFBF}, corresponding to systems without and with ZFBF, respectively. Fig. \ref{DataDiffR} illustrates the channel capacity after ZFBF as a function of the receiver aperture configuration, varying with the number of channels denoted by $N$. In this work, the available transmitted OAM modes $\mathcal{N}$ span from 0 to 5, to match with the FMF-supported modes ranges from $\mathrm{LP}_{01}$ to $\mathrm{LP}_{51}$ \cite{FMF6Modes}. In addition, we make the common assumption for thin lens systems that the aperture diameter $D$ is equal to the focal length $f$, thereby simplifying the optical design and aligning with standard optical system assumptions \cite{AOA}. As we can see, for the system with a single channel, the achievable data rate commences at 0.18 Tb/s, which is in contrast to systems with $N \geq 1$ channels. Moreover, under a small telescope regime, with the increase of aperture size, the data rate usually grows much faster for the systems with smaller $N$. The aforementioned results can be attributed to two main reasons. First, with more concentrated intensity distribution, the single-channel system with incident fundamental Gaussian mode features the most resistance to pointing error, compared with other incident higher-order modes with the donuts-shaped intensity profile in multiple-channel systems. The $\mathrm{LP}_{01}$ fiber mode is more susceptible to angular deviations attributed to the smaller field radius compared with other higher-order $\mathrm{LP}$ modes. However, the channel fluctuations introduced by AOA error can be regarded as a diffracted image jitter on the fiber plane where $d_i = \varepsilon f$, and the performance degradation due to angular deviations can be significantly mitigated in a small aperture regime \cite{AOA}.

On the other hand, the spot radius for the higher-order incident beam mode is positively correlated with the order number. This correlation results in limited power being captured by the small aperture, causing a lag in performance enhancement as the size of the aperture is increased. Moreover, the overall trend initially exhibits an increase followed by a subsequent decrease. This pattern can be attributed to the fact that when the aperture is small, increasing the lens size allows for the collection of more optical power, which in turn makes the system more tolerant to pointing error, as also elucidated in prior research \cite{AOA}. This increased tolerance is due to the larger lens's ability to capture a greater portion of the diffused beam, which is particularly advantageous when precise alignment cannot be maintained, thus enhancing the channel capacity up to a certain point. However, when the aperture is big enough, the resulting larger focal length results in a smaller field of view (FOV), which makes the system more vulnerable to AOA deviation \cite{AOA}. Therefore, the trade-off between aperture diameter $D$ and focal length $f$ results in the optimal telescope configuration to maximize the system performance. Specifically, for $N$ equals 1 to 6, the optimal aperture diameter can be determined as 6 mm, 12 mm, 15 mm, 19 mm, 22 mm, and 24 mm, respectively.

\begin{figure}[t]
   \centering
   \subfloat[]{
        \includegraphics[scale=0.75]{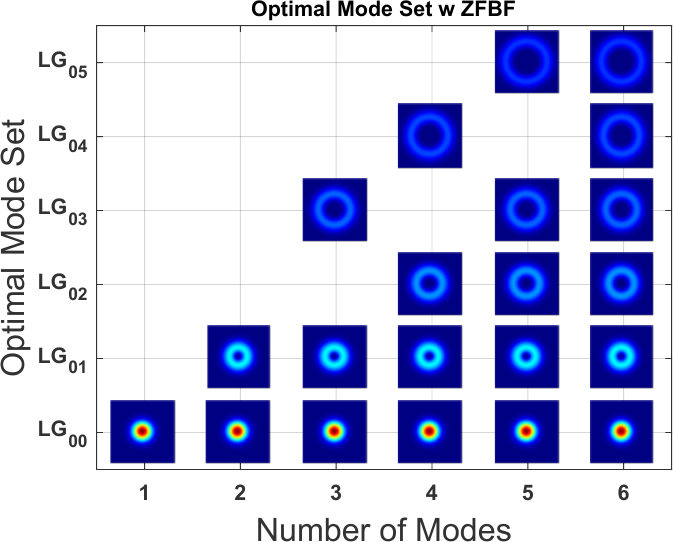}
        \label{ModeSetwZFBF}}
        \\
    \subfloat[]{
        \includegraphics[scale=0.75]{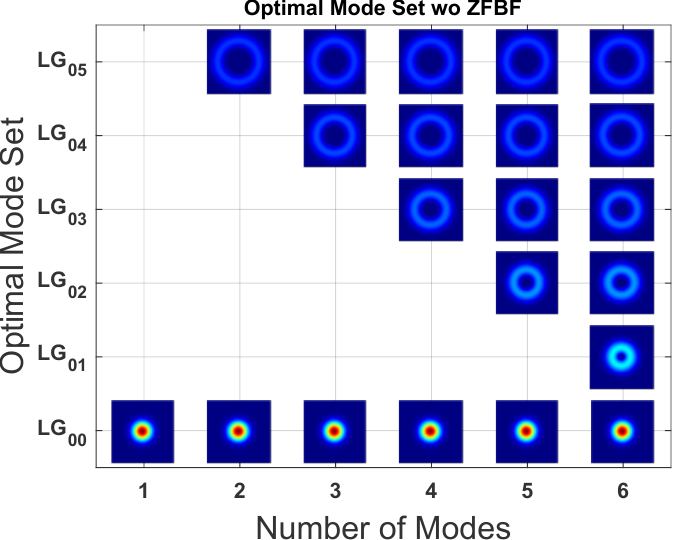}
        \label{ModeSetwoZFBF}}
    \caption{The optimal mode set $\mathcal{N}$ to maximize the channel capacity under different number of multiplexing channels for (a): mutually coherent system without ZFBF, and (b): mutually coherent system with ZFBF. }
    \label{ModesetSelection}
    \vspace{-5mm}
\end{figure}
Notably, in SMM-enabled OWC systems, the choice of the transmitted spatial mode set $\mathcal{N}$ is of paramount importance due to the varying crosstalk attributes exhibited by each mode during misaligned propagation. Fig. \ref{ModesetSelection} depicts the optimal mode selection for both systems with and without ZFBF through exhaustive search. As depicted in Fig. \ref{ModeSetwZFBF}, mode 0 is invariably selected across all considered mode sets 
$\mathcal{N}$ owing to the fundamental Gaussian mode's robust resistance to vibrations as evidenced in Fig. \ref{LG00FixedMisalignment}. For cases where $N \geq 2$, the mode set $\mathcal{N} = \{0, 1\} $ consistently emerges as the preferred choice. This preference is largely due to the effective mitigation of inter-channel interference by ZFBF, which in turn accentuates the importance of signal power when calculating the channel capacity. Additionally, it should be noted that in configurations requiring the transmission of more than a binary set of modes, the selection of adjacent modes results in substantial power consumption for interference compensation. As observed in \eqref{Eq.ZFBFPower}, unlike the utilization of SVD as a linear coding strategy to preserve the signal power, the ZFBF may require substantial power allocation to the transmitted signals to compensate for the interference. Such compensation can lead to the non-uniform and occasionally excessive amplification of power in specific data streams, as it strives to attain the required level of interference mitigation. This outcome brings the trade-off between allocating power for signal transmission and power for compensating interference, an essential consideration in the design and optimization of such communication systems. For example, for $N = 3$, the optimal mode set is $\{0, 1, 3\}$ instead of $\{0, 1, 2\}$, and for $N = 5$,  the optimal mode set is $\{0, 1, 2, 3, 5\}$ instead of $\{0, 1, 2, 3, 4\}$. 

On the other hand, as demonstrated in Fig. \ref{ModeSetwoZFBF}, the absence of ZFBF significantly alters the mode selection strategy, where the importance of maximizing the relative separations between transmitted mode states to mitigate modal crosstalk has been underscored as elucidated in previous works \cite{OAMDis,ShenjieSMMDiversity}. This shift can be attributed to the fact that power is more likely to leak into adjacent modes as discussed in Fig. \ref{SimCouplingEfficiency}, and the presence of inter-channel interference accentuates the importance of mitigating interference power, which is in contrast with systems where ZFBF is employed. Additionally, in configurations where transmission involves more than a binary set of modes, it has been observed that there is a tendency for power from higher-order modes to preferentially couple into modes that are more distantly located within the spectrum. This behavior calls for a deliberate strategy in selecting modes that are adjacent to the highest-ordered mode in the set, to facilitate efficient transmission and to mitigate the effects of power coupling into the unintended modes. For example, when $N= 3$, the optimal mode set is $\mathcal{N} = \{0, 4, 5\} $, and when $N= 4$, the optimal mode set is $\mathcal{N} = \{0, 3, 4, 5\}$.

\begin{figure}[t]
   \centering
    \includegraphics[scale=0.63]{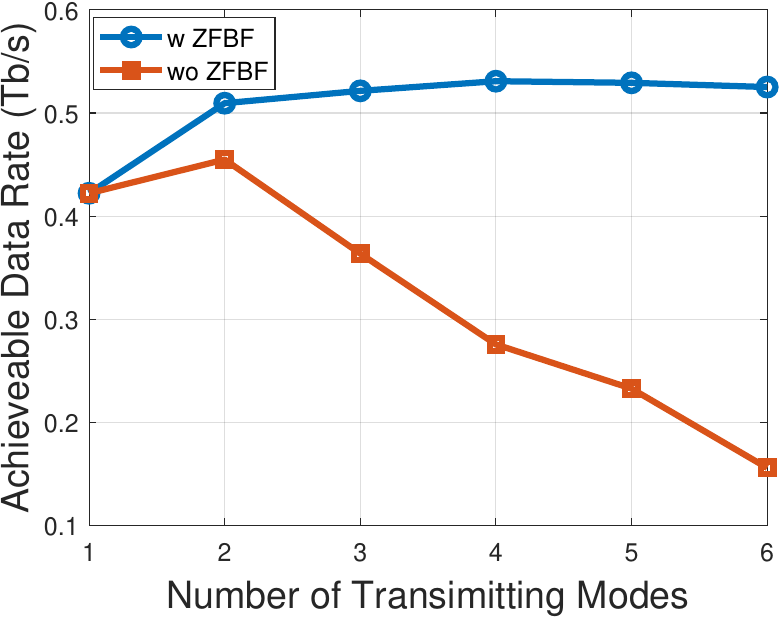}
    \caption{The optimal mode set to maximize the channel capacity under different numbers of multiplexing channels $N$}
    \label{DataRateModeNumber}
\end{figure}

Fig. \ref{DataRateModeNumber} illustrates the aggregated achievable data rate as the function of the number of subchannels for SMM systems with and without ZFBF. Note that the optimal aperture configurations and mode sets shown in Fig. \ref{DataDiffR} and Fig. \ref{ModesetSelection}, respectively, are utilized to attain this result.  It is observable that as the size of the available mode set $N$ expands, the channel capacity with ZFBF exhibits an initial increase followed by a subsequent decrease. This pattern can be attributed to the fact that, with a constrained array of subchannels, ZFBF expends only a minor fraction of the total power to counteract interference. Consequently, the system is able to benefit from the augmented spatial degree of freedom, leading to a marked elevation in the channel capacity. However, adding more transmitted modes significantly decreases the allocated power to each channel under the constant power constraint. On the other hand, more subchannels require increased power to mitigate the crosstalk between each mode. As a result, when the number of modes $N$ is increased, only limited power can be applied to the information-bearing signals, which in turn results in a decreased received power for each channel, leading to a significant drop in the achievable data rate. Therefore, a trade-off emerges between the power used for interference elimination and that used for signal transmission. This leads to the identification of an optimal mode number for the ZFBF-based system as illustrated in Fig. \ref{DataRateModeNumber}, which is realized when $N = 4$, corresponding to the mode set $\mathcal{N} = \{0, 1, 2, 4\}$. Based on the previous discussion, the transmission of different OAM modes through the misaligned channels is known to exhibit variability in maintaining the transmitted power within the original mode state. Concurrently, the degree of power leakage from one channel to another is dependent on the specific states of the modes involved. This leads to an inconsistency in the average signal and interference power across channels, resulting in the multiplexed channels within the SMM system with inherently non-uniform characteristics.

Regarding system performance in the absence of ZFBF, the selection of higher-order OAM modes results in a pronounced increase in crosstalk under the real-world vibration parameter. Coupled with the uniform distribution of signal power across all transmission modes, there is a consequential reduction in the SINR within individual subchannels, which in turn precipitates a decline in system performance when $N \geq 3$. Consequently, the data presented in Fig. \ref{DataRateModeNumber} show that implementing ZFBF can facilitate a significant enhancement in system performance under, namely,  43\% improvement when $N = 3$ and 236\% improvement when $N = 6$. Notably, the findings presented in Fig. \ref{DataRateModeNumber} hold significant implications for the design of practical OWC-based SMM systems. The limitations set by the physical dimensions of transceivers in actual deployments are crucial factors that restrict the operational range, thereby highlighting the necessity of choosing the optimal combination of modes to realize the maximum channel capacity within these constraints.

\begin{figure}[t]
\includegraphics[scale=0.6]{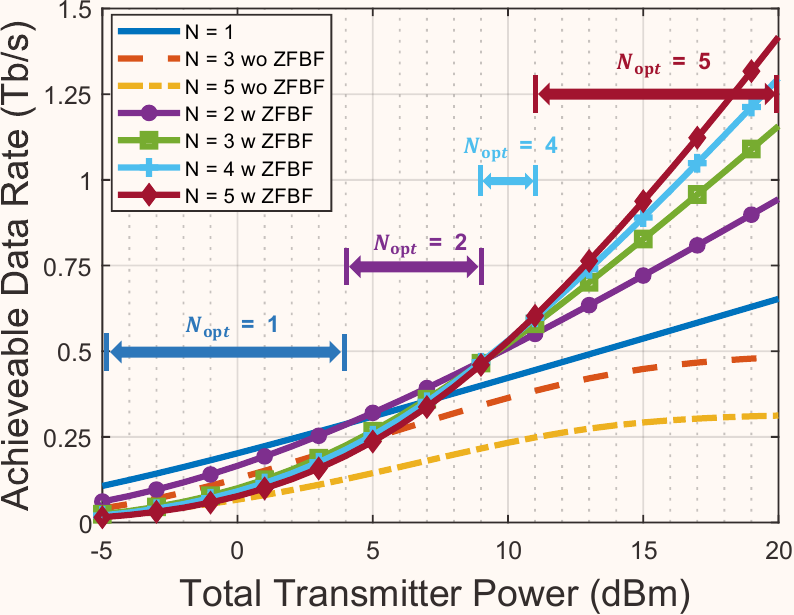}

\caption{ The achievable data rate with various $N$ under different transmission power $\xi_t$ (from -15 dBm to 30 dBm), with arrows indicating the optimal mode number under corresponding power budgets.}
\label{DataRateTransimittedPower}
\end{figure}
The achievable data rate with respect to $\xi_t$ for various $N$ under the optimal mode set is illustrated in Fig. \ref{DataRateTransimittedPower}. The depicted trends reveal that, for the ZFBF-based system operating under low-power conditions, a configuration with a smaller number of subchannels can exhibit superior performance relative to a system with a larger $N$. Specifically, when the transmission power falls below 4 dBm, the single-channel system surpasses the performance of all multiplexed configurations, and the 5-mode system features the best performance as the $\xi_t$ is larger than 11 dBm. This is because under the power-limited scenario, as more power is used to compensate for the interference among subchannels when increasing $N$, less signal power will result in the degradation in channel capacity as the constant noise variance. However, under the high-power situation, since more power can be applied to the information-bearing signal, the SNR can experience a huge enhancement under such a constant noise power. Therefore, with the larger power budget, the larger $N$ can benefit more from the extra spatial DOF with the elimination of interference. For example, with $\xi_t$ is 11 dBm, the 5-channel can feature 35.6$\%$ and 4.8$\%$ improvement when compared with the single-channel and 3-channel system, while the enhancement can be increased to 117.2$\%$ and 22.5$\%$ as the power budget is extended to 20 dBm. Hence, the selection of a specific number of subchannels should be tailored to different power budgets. As concluded by the arrows, after applying the ZFBF, $N = 1$ is the optimal option when the transmission power is less than 4 dBm. As further expanding the power budget, a system with two subchannels outperforms others up to a transmission power $\xi_t \leq 9$ dBm. In addition, as the power ranges between 9 dBm and 11 dBm, $N = 4$ maximizes the system capacity. Finally, for a power budget exceeding 11 dBm, a configuration with five subchannels ($N = 5$) optimizes system performance. On the other hand, for the system without ZFBF, one can see that the performance shows a tendency toward saturation in high-power regimes. This saturation occurs because an increase in transmission power concurrently amplifies the interference power. Therefore, as evidenced by Fig. \ref{DataRateTransimittedPower}, the ZFBF not only allows the system to capitalize on the spatial DOF but also facilitates the system performance enhancement through the escalation of transmission power.

\section{Conclusion}
\label{Conclusion}
In this paper, we investigate the SMM-enabled FMF coupling system based on the IM/DD receiver with mutually coherent channels impaired by link misalignment. We introduce a theoretical method that provides more accurate and compact misalignment-impaired channel models compared with the prior work in the literature. Building upon this, we develop a theoretical coupling efficiency model from the multi-mode OWC link to the FMF. It is found that in the presence of misalignment, the coupling efficiency varies significantly for the transmission of different modes. Our research also observes significant intermodal power leakage due to device vibrations, for which we employ the ZFBF to mitigate subchannel crosstalk and accordingly derive the analytical aggregated data rate equations. Through simulations, we identify optimal system configurations that maximize performance, including aperture design, number of transmitting modes, and mode set selection. These optimal configurations result in a performance enhancement of over 200\% when using ZFBF. Considering the system performance is significantly altered by the received power, this paper finally presents the most effective mode exploitation strategies for varying power budgets. 
\bibliographystyle{ieeetr}
\bibliography{IEEEexample}

\begin{thebibliography}{10}

\bibitem{OWCOverall}
K.~Wang, T.~Song, Y.~Wang, C.~Fang, J.~He, A.~Nirmalathas, C.~Lim, E.~Wong, and S.~Kandeepan, ``Evolution of short-range optical wireless communications,'' {\em Journal of Lightwave Technology}, vol.~41, no.~4, pp.~1019--1040, 2023.

\bibitem{OWCsecurity}
X.~Zhang, G.~Klevering, X.~Lei, Y.~Hu, L.~Xiao, and G.-h. Tu, ``The security in optical wireless communication: A survey,'' {\em ACM Computing Surveys}, 2023.

\bibitem{FiberCouplingDemo3}
Y.~Hong, F.~Feng, K.~R.~H. Bottrill, N.~Taengnoi, R.~Singh, G.~Faulkner, D.~C. O'Brien, and P.~Petropoulos, ``Demonstration of \&gt;1tbit/s wdm owc with wavelength-transparent beam tracking-and-steering capability,'' {\em Opt. Express}, vol.~29, pp.~33694--33702, Oct 2021.

\bibitem{FiberCouplingDemo2}
A.~Schreier, O.~Alia, R.~Wang, R.~Singh, G.~Faulkner, G.~Kanellos, R.~Nejabati, D.~Simeonidou, J.~Rarity, and D.~O'Brien, ``Coexistence of quantum and 1.6 tbit/s classical data over fibre-wireless-fibre terminals,'' {\em Journal of Lightwave Technology}, vol.~41, no.~16, pp.~5226--5232, 2023.

\bibitem{FiberCouplingDemo1}
R.~Singh, F.~Feng, Y.~Hong, G.~Faulkner, R.~Deshmukh, G.~Vercasson, O.~Bouchet, P.~Petropoulos, and D.~O’Brien, ``Design and characterisation of terabit/s capable compact localisation and beam-steering terminals for fiber-wireless-fiber links,'' {\em Journal of Lightwave Technology}, vol.~38, no.~24, pp.~6817--6826, 2020.

\bibitem{OWDCNfeasibily}
S.~Zhang, X.~Xue, F.~Yan, B.~Pan, X.~Guo, K.~Mekonnen, E.~Tangdiongga, and N.~Calabretta, ``Feasibility study of optical wireless technology in data center network,'' {\em IEEE Photonics Technology Letters}, vol.~33, no.~15, pp.~773--776, 2021.

\bibitem{OWCDCN}
S.~Zhang, R.~Kraemer, B.~Pan, X.~Xue, K.~Prifti, F.~Yan, X.~Guo, E.~Tangdiongga, and N.~Calabretta, ``Experimental assessment of a novel optical wireless data center network architecture,'' in {\em 2020 European Conference on Optical Communications (ECOC)}, pp.~1--4, 2020.

\bibitem{OWDCNSurvey}
A.~S. Hamza, J.~S. Deogun, and D.~R. Alexander, ``Wireless communication in data centers: A survey,'' {\em IEEE communications surveys \& tutorials}, vol.~18, no.~3, pp.~1572--1595, 2016.

\bibitem{PON}
K.~Pang, H.~Song, X.~Su, K.~Zou, Z.~Zhao, H.~Song, A.~Almaiman, R.~Zhang, C.~Liu, N.~Hu, S.~Zach, N.~Cohen, B.~Lynn, A.~F. Molisch, R.~W. Boyd, M.~Tur, and A.~E. Willner, ``Experimental mitigation of the effects of the limited size aperture or misalignment by singular-value-decomposition-based beam orthogonalization in a free-space optical link using laguerre--gaussian modes,'' {\em Opt. Lett.}, vol.~45, pp.~6310--6313, Nov 2020.

\bibitem{OWPON}
N.~Ghazisaidi, M.~Scheutzow, and M.~Maier, ``Survivability analysis of next-generation passive optical networks and fiber-wireless access networks,'' {\em IEEE Transactions on Reliability}, vol.~60, no.~2, pp.~479--492, 2011.

\bibitem{photoniclanternoverview}
N.~K. Fontaine, J.~Carpenter, S.~Gross, S.~Leon-Saval, Y.~Jung, D.~J. Richardson, and R.~Amezcua-Correa, ``Photonic lanterns, 3-d waveguides, multiplane light conversion, and other components that enable space-division multiplexing,'' {\em Proceedings of the IEEE}, vol.~110, no.~11, pp.~1821--1834, 2022.

\bibitem{TaperedPhotonicLattern}
W.~Guo, Y.~Li, J.~Chen, T.~Jin, S.~Jiao, J.~Wu, J.~Qiu, and H.~Guo, ``Satellite-to-ground optical downlink model using mode mismatching multi-mode photonic lanterns,'' {\em Opt. Express}, vol.~31, pp.~35041--35053, Oct 2023.

\bibitem{ULIPhotonicLattern}
D.~G. MacLachlan, R.~J. Harris, D.~Choudhury, R.~D. Simmonds, P.~S. Salter, M.~J. Booth, J.~R. Allington-Smith, and R.~R. Thomson, ``Development of integrated mode reformatting components for diffraction-limited spectroscopy,'' {\em Opt. Lett.}, vol.~41, pp.~76--79, Jan 2016.

\bibitem{PhotonicIntegratedCombiner}
X.~Ji, J.~Liu, J.~He, R.~N. Wang, Z.~Qiu, J.~Riemensberger, and T.~J. Kippenberg, ``Compact, spatial-mode-interaction-free, ultralow-loss, nonlinear photonic integrated circuits,'' {\em Commun. Phys.}, vol.~5, no.~1, p.~84, 2022.

\bibitem{TurblenceFMF1}
F.~Wang, C.~Qiu, Y.~Chen, and G.~Hu, ``Performance of improved mode diversity reception for free-space optical communication under atmospheric turbulence,'' {\em J. Opt. Commun. Netw.}, vol.~14, pp.~725--732, Sep 2022.

\bibitem{FMFCouplingDis}
X.~Fan, D.~Wang, J.~Cheng, J.~Yang, and J.~Ma, ``Few-mode fiber coupling efficiency for free-space optical communication,'' {\em Journal of Lightwave Technology}, vol.~39, no.~6, pp.~1823--1829, 2021.

\bibitem{CapacityOAMMIMO}
N.~Zhao, X.~Li, G.~Li, and J.~M. Kahn, ``Capacity limits of spatially multiplexed free-space communication,'' {\em Nature photonics}, vol.~9, no.~12, pp.~822--826, 2015.

\bibitem{ShenjieSMMDiversity}
S.~Huang, G.~R. Mehrpoor, and M.~Safari, ``Spatial-mode diversity and multiplexing for fso communication with direct detection,'' {\em IEEE Transactions on Communications}, vol.~66, no.~5, pp.~2079--2092, 2018.

\bibitem{OAM1}
A.~E. Willner, H.~Song, K.~Zou, H.~Zhou, and X.~Su, ``Orbital angular momentum beams for high-capacity communications,'' {\em J. Lightwave Technol.}, vol.~41, pp.~1918--1933, Apr 2023.

\bibitem{OAMDis}
G.~Xie, L.~Li, Y.~Ren, H.~Huang, Y.~Yan, N.~Ahmed, Z.~Zhao, M.~P.~J. Lavery, N.~Ashrafi, S.~Ashrafi, R.~Bock, M.~Tur, A.~F. Molisch, and A.~E. Willner, ``Performance metrics and design considerations for a free-space optical orbital-angular-momentum multiplexed communication link,'' {\em Optica}, vol.~2, pp.~357--365, Apr 2015.

\bibitem{Buildingsway}
S.~Arnon, ``Effects of atmospheric turbulence and building sway on optical wireless-communication systems,'' {\em Opt. Lett.}, vol.~28, pp.~129--131, Jan 2003.

\bibitem{Datacentervibration}
M.~Curran, K.~Zheng, H.~Gupta, and J.~Longtin, ``Handling rack vibrations in fso-based data center architectures,'' in {\em 2018 International Conference on Optical Network Design and Modeling (ONDM)}, pp.~47--52, IEEE, 2018.

\bibitem{PointingError}
A.~A. Farid and S.~Hranilovic, ``Outage capacity optimization for free-space optical links with pointing errors,'' {\em Journal of Lightwave Technology}, vol.~25, no.~7, pp.~1702--1710, 2007.

\bibitem{AOA}
S.~Huang and M.~Safari, ``Free-space optical communication impaired by angular fluctuations,'' {\em IEEE Transactions on Wireless Communications}, vol.~16, no.~11, pp.~7475--7487, 2017.

\bibitem{LGBEAMDis}
K.~Pang, H.~Song, X.~Su, K.~Zou, Z.~Zhao, H.~Song, A.~Almaiman, R.~Zhang, C.~Liu, N.~Hu, S.~Zach, N.~Cohen, B.~Lynn, A.~F. Molisch, R.~W. Boyd, M.~Tur, and A.~E. Willner, ``Experimental mitigation of the effects of the limited size aperture or misalignment by singular-value-decomposition-based beam orthogonalization in a free-space optical link using laguerre--gaussian modes,'' {\em Opt. Lett.}, vol.~45, pp.~6310--6313, Nov 2020.

\bibitem{MIMOfIBER}
I.~Gasulla and J.~M. Kahn, ``Performance of direct-detection mode-group-division multiplexing using fused fiber couplers,'' {\em Journal of Lightwave Technology}, vol.~33, no.~9, pp.~1748--1760, 2015.

\bibitem{MIMODSPFreeSpace1}
Y.~Li, Z.~Hu, D.~M. Benton, A.~Ali, M.~Patel, and A.~D. Ellis, ``Demonstration of 10-channel mode-and polarization-division multiplexed free-space optical transmission with successive interference cancellation dsp,'' {\em Optics Letters}, vol.~47, no.~11, pp.~2742--2745, 2022.

\bibitem{MIMODSPFreeSpace2}
Y.~Li, Z.~Chen, Z.~Hu, D.~M. Benton, A.~A.~I. Ali, M.~Patel, M.~P.~J. Lavery, and A.~D. Ellis, ``Enhanced atmospheric turbulence resiliency with successive interference cancellation dsp in mode division multiplexing free-space optical links,'' {\em Journal of Lightwave Technology}, vol.~40, no.~24, pp.~7769--7778, 2022.

\bibitem{MutuallyIncoherentSVD}
K.~Pang, H.~Song, X.~Su, K.~Zou, Z.~Zhao, H.~Song, A.~Almaiman, R.~Zhang, C.~Liu, N.~Hu, {\em et~al.}, ``Experimental mitigation of the effects of the limited size aperture or misalignment by singular-value-decomposition-based beam orthogonalization in a free-space optical link using laguerre--gaussian modes,'' {\em Optics Letters}, vol.~45, no.~22, pp.~6310--6313, 2020.

\bibitem{MutuallyInCoandCo}
Y.~Yadin and M.~Orenstein, ``Parallel optical interconnects over multimode waveguides,'' {\em Journal of Lightwave Technology}, vol.~24, no.~1, pp.~380--386, 2006.

\bibitem{MutuallyInDiffLaser}
H.~R. Stuart, ``Dispersive multiplexing in multimode optical fiber,'' {\em Science}, vol.~289, no.~5477, pp.~281--283, 2000.

\bibitem{MutuallyCohernet}
Y.~Yadin and M.~Orenstein, ``Parallel optical interconnects over multimode waveguides using mutually coherent channels and direct detection,'' {\em Journal of lightwave technology}, vol.~25, no.~10, pp.~3126--3131, 2007.

\bibitem{ShenjieSMMZFBF}
S.~Huang and M.~Safari, ``Spatial-mode multiplexing with zero-forcing beamforming in free space optical communications,'' in {\em 2017 IEEE International Conference on Communications Workshops (ICC Workshops)}, pp.~331--336, IEEE, 2017.

\bibitem{Phaseretravel1}
K.~Choutagunta, I.~Roberts, D.~A. Miller, and J.~M. Kahn, ``Adapting mach--zehnder mesh equalizers in direct-detection mode-division-multiplexed links,'' {\em Journal of Lightwave Technology}, vol.~38, no.~4, pp.~723--735, 2019.

\bibitem{OptimizingFMFEfficiency}
A.~Fardoost, H.~Wen, H.~Liu, F.~G. Vanani, and G.~Li, ``Optimizing free space to few-modefiber coupling efficiency,'' {\em Applied Optics}, vol.~58, no.~13, pp.~D34--D38, 2019.

\bibitem{FOVDataRateTradeoff}
M.~D. Soltani, H.~Kazemi, E.~Sarbazi, T.~E. El-Gorashi, J.~M. Elmirghani, R.~V. Penty, I.~H. White, H.~Haas, and M.~Safari, ``High-speed imaging receiver design for 6g optical wireless communications: a rate-fov trade-off,'' {\em IEEE Transactions on Communications}, vol.~71, no.~2, pp.~1024--1043, 2022.

\bibitem{SMFCouplingEfficiency1}
P.~J. Winzer and W.~R. Leeb, ``Fiber coupling efficiency for random light and its applications to lidar,'' {\em Optics letters}, vol.~23, no.~13, pp.~986--988, 1998.

\bibitem{SMFCouplingEfficiency2}
Y.~Dikmelik and F.~M. Davidson, ``Fiber-coupling efficiency for free-space optical communication through atmospheric turbulence,'' {\em Applied Optics}, vol.~44, no.~23, pp.~4946--4952, 2005.

\bibitem{moderelation}
R.~Br{\"u}ning, Y.~Zhang, M.~McLaren, M.~Duparr{\'e}, and A.~Forbes, ``Overlap relation between free-space laguerre gaussian modes and step-index fiber modes,'' {\em JOSA A}, vol.~32, no.~9, pp.~1678--1682, 2015.

\bibitem{PhotonicsBook}
B.~E. Saleh and M.~C. Teich, {\em Fundamentals of photonics}.
\newblock john Wiley \& sons, 2019.

\bibitem{SlantLGBeams}
M.~V. Vasnetsov, V.~A. Pas'ko, and M.~S. Soskin, ``Analysis of orbital angular momentum of a misaligned optical beam,'' {\em New Journal of Physics}, vol.~7, p.~46, feb 2005.

\bibitem{Nearfield}
S.~Sun, N.~An, F.~Yang, J.~Song, and Z.~Han, ``Capacity characterization analysis of optical intelligent reflecting surface assisted miso vlc,'' {\em IEEE Internet of Things Journal}, vol.~11, no.~3, pp.~4801--4814, 2024.

\bibitem{IMDDDataRate}
J.-B. Wang, Q.-S. Hu, J.~Wang, M.~Chen, and J.-Y. Wang, ``Tight bounds on channel capacity for dimmable visible light communications,'' {\em Journal of Lightwave Technology}, vol.~31, no.~23, pp.~3771--3779, 2013.

\bibitem{Eyesafety}
M.~Dehghani~Soltani, E.~Sarbazi, N.~Bamiedakis, P.~d. Souza, H.~Kazemi, J.~M.~H. Elmirghani, I.~H. White, R.~V. Penty, H.~Haas, and M.~Safari, ``Safety analysis for laser-based optical wireless communications: A tutorial,'' {\em Proceedings of the IEEE}, vol.~110, no.~8, pp.~1045--1072, 2022.

\bibitem{FMF6Modes}
A.~Wang and L.~Zhu, ``Deep learning based mode group recognition for mode division multiplexing in conventional multimode fiber,'' in {\em 2019 Asia Communications and Photonics Conference (ACP)}, pp.~1--3, 2019.

\end{thebibliography}

\vfill

\end{document}